\documentclass[twocolumn,prb]{revtex4} 
\usepackage{graphicx}

\newcommand{\BEQ}{\begin{equation}}
\newcommand{\EEQ}{\end{equation}}
\newcommand{\BEA}{\begin{eqnarray}}
\newcommand{\EEA}{\end{eqnarray}}

\newcommand{\D}{{\cal {D}}}
\newcommand{\h}{{\tilde h}}
\renewcommand{\d}{{\rm d }}

\newcommand{\dx}{{\rm d}x\,}
\newcommand{\dy}{{\rm d}y\,}

\newcommand{\p}{\partial}

\newcommand{\nn}{\nonumber }
\newcommand{\Kt}{{\tilde K}}
\newcommand{\Tr}{{\rm Tr}}

\newcommand{\s}{{\sigma}}

\newcommand{\eab}{\epsilon_{ab}}
\newcommand{\ebc}{\epsilon_{bc}}
\newcommand{\eca}{\epsilon_{ca}}
\newcommand{\eac}{\epsilon_{ac}}

\newcommand{\ecd}{\epsilon_{cd}}
\newcommand{\ede}{\epsilon_{de}}
\newcommand{\eae}{\epsilon_{ae}}
\newcommand{\ead}{\epsilon_{ad}}
\newcommand{\eda}{\epsilon_{da}}

\begin{document}
\title{Thermodynamic Properties and Phase Transitions
in a Mean-Field Ising Spin Glass on Lattice Gas: the Random 
Blume-Emery-Griffiths-Capel Model }

\date{\today}
\author{Andrea Crisanti and Luca Leuzzi}
\affiliation{Department of Physics and SMC INFM Center, \\
University of Rome I,
``La Sapienza'',\\
 Piazzale A. Moro 2, 00185, Rome, Italy}
\begin{abstract}
The study of the mean-field static solution of the Random
Blume-Emery-Griffiths-Capel model, an Ising-spin lattice gas with
quenched random magnetic interaction, is performed.  The model
exhibits a paramagnetic phase, described by a stable Replica Symmetric
solution.  When the temperature is decreased or the density increases,
the system undergoes a phase transition to a Full Replica Symmetry
Breaking spin-glass phase.  The nature of the transition can be either
of the second order (like in the Sherrington-Kirkpatrick model) or, at
temperature below a given critical value, of the first order in the
Ehrenfest sense, with a discontinuous jump of the order parameter and
accompanied by a latent heat.  In this last case coexistence of phases
takes place.  The thermodynamics is worked out in the Full Replica
Symmetry Breaking scheme, and the relative Parisi equations are solved
using a pseudo-spectral method down to zero temperature.
\end{abstract}
\maketitle 

%%%%%%%%%%%%%%%%%%%%%%%%%%%%%%%%%%%%%%%%%%%%%%%%%%%%%%%

\label{c:cap7}

Since its discovery, the spin glass (SG)
phase  has played and still plays a  fundamental role
in the  investigation  and understanding of many basic properties of  
disordered and complex systems.
The analysis of the mean-field approximation of theoretical models
displaying such a phase has revealed different possible scenarios,
including different kinds of transition from the paramagnetic phase to
the SG phase, as well as different kinds of SG phases.  Most of the
work, however, has been concentrated on just two scenarios.

In order of appearance in literature the first scenario is described
by a Full Replica Symmetry Breaking (FRSB) solution characterized by a
continuous order parameter function, \cite{PJPA80} which continuously
grows from zero by crossing the transition.  The prototype model is
the Sherrington-Kirkpatrick (SK) model, \cite{SKPRL75} a fully
connected Ising-spin model with quenched random magnetic interactions.

The second scenario, initially introduced by Derrida by means of the
Random Energy Model (REM),\cite{DPRL80} provides a transition with a
jump in the order parameter to a stable low temperature phase in which
the replica symmetry is spontaneously broken only once.  The order
parameter is a step function taking two values $q_{\rm min}$ and the
so-called Edwards-Anderson order parameter, \cite{EAJPF75} $q_{\rm
EA}$, (else said {\em self-overlap}), with $q_{\rm min}<q_{\rm EA}$.
In the paramagnetic phase they are both equal to zero.  At the
transition, $q_{\rm EA}$ grows to a value larger than $q_{\rm min}$.
No discontinuity appear, however, in the thermodynamic functions.
 Actually, at the transition to the one step Replica Symmetry Breaking
(1RSB) SG phase, the Edwards-Anderson order parameter $q_{EA}$ can
either grow continuously from zero or jump discontinuously to a finite
value.  The first case of this second scenario includes Potts-glasses
with three or four states, \cite{GKSPRL85} the spherical $p$-spin
spin-glass model in strong magnetic field \cite{CSZPB92} and some {\em
inhomogeneous} spherical $p$-spin model with a mixture of $p=2$ and
$p>3$ interactions.\cite{Nie95,CriCuk}  The latter case includes,
instead, Potts-glasses with more than four states \cite{GKSPRL85},
quadrupolar glass models, \cite{GKSPRL85,GolShe85} $p$-spin
interaction spin-glass models with $p>2$
\cite{DPRL80,GMNPB84,GNPB85,KWPRA87} and the spherical $p$-spin
spin-glass model in weak magnetic field. \cite{CSZPB92} 
Because of the discontinuity of the overlap parameter
across the transition the models
belonging to this second case, often referred to as ``discontinuous
spin glasses''  (see, for instance, Ref. [\onlinecite{BCKM98}]), have
been widely investigated in the last years because of their relevance
for the structural glass transition.
\cite{KWPRA87,CHSZPB93}

In all scenarios discussed above no latent heat occurs, i.e. the phase
transition  is continuous in the Ehrenfest sense.  In this
paper, instead, we consider a spin glass model undergoing (below a
given critical point) a true thermodynamic first order phase
transition between a paramagnetic (PM) and a Full Replica Symmetry
Breaking (FRSB) SG phase, presenting coexistence of phases and latent
heat, completing the work presented in Ref. [\onlinecite{CLPRL02}].

\begin{figure}[t!]
\begin{center}
{\includegraphics*[width=.49\textwidth]{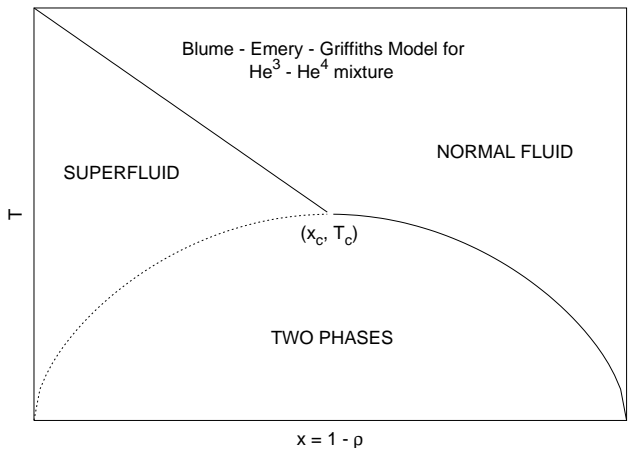}}
\end{center}
\protect\caption{ \small{The phase diagram of the mean field model for
He$^3$-He$^4$ mixture as originally studied by BEG. $x$ is the density
of He$^3$ particles and $\rho$ the density of He$^4$.  }}
\label{fig:BEG}
\end{figure}

Such a model is a generalization of the Blume-Emery-Griffiths-Capel
(BEGC) model \cite{BEGPRA71,HBPRL91} for the $\lambda$ transition and
the phase separation in the mixtures of He$^3$-He$^4$ in a crystal
field, in which a quenched disordered magnetic interaction is
introduced.  In that case the phase diagram consisted of a fluid
phase, a super-fluid one and a mixed phase (see Fig. \ref{fig:BEG}
for a pictorial representation of the diagram).

The first study of a spin-glass model undergoing a genuine first order
thermodynamic transition is the Ghatak-Sherrington model (GS),
\cite{GSJPC77} a simplified version of the model under current
investigation.  In this spin-$1$ model in crystal field, no
biquadratic coupling is considered: the Replica Symmetric (RS)
solution and its stability have been carried out by Lage and Almeida,
\cite{LAJPC82} Mottishaw and Sherrington \cite{MSJPC85} and da Costa
{\em { et al.}}\cite{dCYSJPA94}, albeit not always with compatible
results.  There the evidence for a first order phase transition was
found, in the neighborhood of the tricritical point.

In the last few years some work has been devoted to the comprehension
of the Random generalization of the BEGC model (RBEGC), mostly at the
level of the RS solution, that turns out to be unstable as soon as the
transition takes place.  Such an analysis has been performed in
Refs. [\onlinecite{ANSJPF96,SNAJPF97}] for what concerns the lattice gas
version of the model, and in
Refs. [\onlinecite{dCNYJPA97,SEPJB99,dCAEPJB00,ANdCJPC00}] for the spin-$1$
model.  For sake of completeness we also report the study
[\onlinecite{AdCNEPJB00}] where, together with the random magnetic
interaction, also a random biquadratic interaction is considered.

Quite recently, the spherical version of the present model has been worked out
exhaustively for positive and null particle-particle interaction, 
both statically and dynamically, by Caiazzo et. {\em al.}\cite{CNC1,CNC2}.
 
In section \ref{s:rbeg} we present the model, also indicating the
connection with different notation in the literature. In
Sec. \ref{s:rep} the replica formalism for the model is recalled, the
Replica Symmetric (RS) solution presented, together with its stability
and low temperature analysis.  In Sec. \ref{s:var} we adapt the
variational method for the FRSB solution \cite{SDJPC84} to the model,
we explain the numerical method of resolution applied and we discuss
the practical necessity of using different gauges to express the
thermodynamic functional in different parameters regions. In
Sec. \ref{ss:BEGCTD} we study the thermodynamic observables. From the
behavior of the entropy versus temperature at fixed chemical potential
we observe that below the tricritical point the transition involves
latent heat. In Sec. \ref{ss:PhDi} we display the phase diagrams in the 
parameters $T$, $\mu$ and density.
Sec.  \ref{ss:sum7} contains our conclusions.

	\section{The  {\em{Random Blume-Emery-Griffiths-Capel}} Model}
	\label{s:rbeg}

There exist two completely equivalent versions of  this disordered
model (at least as far as the statics is concerned).
One version is a direct generalization of the BEGC model, 
with spin 1 variables
($\sigma_i=1,0,-1$ on site $i$),\cite{dCNYJPA97}
 the other one is a lattice gas of
Ising spins (spin $S_i= 1,-1$, with occupation numbers
$n_i=0,1$ on site $i$).\cite{ANSJPF96,SNAJPF97}

In this paper we will use the second formulation
described, in the mean-field approximation,
 by the Hamiltonian  \cite{SNAJPF97}
\begin{equation}
\mathcal{H}= - \sum_{i<j} J_{ij} S_i S_j n_i n_j 
  -\frac{K}{N} \sum_{i<j} n_i n_j
    -\mu \sum_i n_i - h \sum_i S_i n_i ,
\label{f:61a}
\end{equation}
where the  Ising spin glass lattice gas in an external magnetic  field 
is coupled to a spin reservoir by the chemical potential $\mu$. 
$K$ is the particle-particle coupling constant and 
$h$ the external magnetic field.
The magnetic interaction is described by 
 quenched Gaussian random variables  $J_{ij}$, symmetric in $i \leftrightarrow
 j$,
with mean ${\overline{J_{ij}}}=J_0/N$ and variance
${\overline{J_{ij}^2}}={\overline{J_{ji}^2}}=J^2/N$. Here and in the 
following  the overline denotes average with respect to disorder.

Just for completeness we also report the Hamiltonian in the original
formulation:
 \begin{equation}
\mathcal{H}= - \sum_{i<j} J_{ij} \s_i \s_j 
  -\frac{K}{N} \sum_{i<j} (\s_i \s_j)^2
    +D \sum_i \s_i^2 - h \sum_i \s_i ,
\nn
\end{equation}
The transformation $\sigma_i=S_i~n_i$, between the spin lattice gas 
dynamic variables and 
the spin-1  variables $\sigma_i$, and the transformation 
$D=-\mu-T \log 2$, 
between $\mu$  and the 
crystal field $D$ of the spin-$1$ system, allow for a perfect equivalence 
of two versions of the  model.

Some limiting cases of the model are the SK model \cite{SKPRL75} (for
$\mu/J\to \infty$)
and the site frustrated percolation model \cite{CJPIV93}
 (for $K=-1$ and $J/\mu\to \infty$).

In most cases studied up to now 
\cite{GSJPC77,LAJPC82,MSJPC85,dCYSJPA94,
ANSJPF96,SNAJPF97,dCNYJPA97,SEPJB99,dCAEPJB00,ANdCJPC00} the analysis
was mainly limited to the RS solution.  The general picture which
emerged was an instability of the RS solution below some transition
line in the region of low temperature and large density of particles.
The nature of the SG phase (1RSB of FRSB) and the type of transition
(SK-like~\cite{PJPA80}, $p$-spin-like \cite{GNPB85} or
different~\cite{MSJPC85}) were, however, not clear, also due to some
``anomalous'' properties at the RS level such as, for instance,
complex stability eigenvalues.  Despite this fact, the possibility of
a $p$-spin-like transition has put new interest into this model as a
possible, more realistic, model for the structural glasses, and its
finite dimensions version has been numerically investigated in a
search for evidence of a structural glass transition scenario
\cite{dCCPRE01,NC}.  Performing the quenched averages in the most general
Replica scheme of computation it has been possible to show that the
stable solution in the mean-field case, is FRSB everywhere in the
 SG phase, where spins are {\rm frozen}. \cite{CLPRL02}

%%%%%%%%%%%%%%%%%%%%%%%%%%%%%%%%%%%%%%%%%%%%%%%%%%%%%%%

	\section{The Replica Trick for the Thermodynamics of 
		Disordered Systems}
	\label{s:rep}
For any fixed (quenched) 
coupling realization $J$, the partition function of a system
of $N$ dynamic (annealed) variables $\sigma$,
 is given by \cite{MPV87,Fischer91}
\begin{equation}
   Z_N[J]= \mbox{Tr}_{\sigma}\ \exp\bigl(-\beta {\cal H}[J;\sigma]\bigr)
\end{equation}
and the quenched free energy per spin is
\begin{equation}
f_N= -{1\over N\beta}\ \overline{\,\log Z_N\,}
        = -{1\over N\beta}\ \int d [J]\ P[J]\ \log Z_N[J]
\end{equation}
where $\overline{(\cdots)}$ indicates the average over the couplings
realizations. The thermodynamic limit of the free energy,
$-\lim_{N\to\infty} \log Z_N[J]\ /\ N\beta$
is well defined and  equal to the quenched free energy
$f=\lim_{N \to \infty} f_N$
for almost any coupling realization $J$ (self-average property).

The analytic computation of the quenched free energy, i.e., of the average of
the logarithm of the partition function, is  quite a difficult problem,
even in simple cases as nearest neighbor one dimensional models.
However, since the integer moments of the partition function are easier
to compute, the standard method involves the so called ``replica trick''
by considering the annealed free energy $f(n)$ of $n$ non-interacting
`replicas' of the system, \cite{SKPRL75,MPV87,Fischer91}
\begin{equation}
f(n)= -\lim_{N\to\infty} {1\over N\beta n}\ \log\left[
                                          \overline{(Z_N[J])^n}
                                               \right].
\end{equation}
The quenched free energy of the original system is then recovered as the
continuation of $f(n)$ down to the unphysical limit $n=0$,
\footnote{Indeed, the important aspect of mean field models is that
this calculation can actually be carried out, whereas for short range
disordered models the free energy is beyond the reach of analytical
computation.}
\begin{equation}
   f= -\lim_{N\to \infty}\lim_{n\to 0}
                         {\overline{(Z_N[J])^n} - 1\over N\beta n}
      = \lim_{n\to 0} f(n).
\end{equation}
In the last equality we assumed that the replica limit and the
thermodynamic limit can be exchanged.
This procedure replaces the original interactions in the real space
with couplings among different replicas. 
The interested reader can find a complete and detailed 
presentation of the replica method for disordered statistical mechanical 
systems in Refs. [\onlinecite{MPV87}] and [\onlinecite{Fischer91}].

Applying the replica trick to the computation of the thermodynamic potential
of the random BEGC model we find, in the saddle point approximation
for large $N$, 
\BEA
&&{\overline{Z_N[J]^n}}=\exp\left\{-n N \beta 
f\left[\{\rho\},\{m\},\{q\}\right]
\right\}
\\
&&\beta f\equiv  \frac{\beta \Kt}{2}
\frac{1}{n}\sum_{a=1}^{n}\rho_a^2+\frac{\beta J_0}{2}\frac{1}{n}\sum_{a=1}^n 
m_a^2
\nn
\\
&&\hspace*{2 cm}
+\frac{\left(\beta J\right)^2}{4}\frac{1}{n}\sum_{a\neq b} q_{ab}^2
-\frac{1}{n}\log Z'
\label{f:62a}
\\
&&Z'\equiv\sum_{\{S\},\{n\}}\exp\left\{-\beta H'[\{d\},\{m\},\{q\}]\right\}
\label{f:62b}
\\
&&-\beta H'\left[\{\rho\},\{m\},\{q\}\right]\equiv\sum_{a=1}^n n_a
\left(\beta \Kt\rho_a +\beta \mu\right)
\label{f:62c} \\
&&\ \ +\sum_{a=1}^nS_an_a\left(\beta J_0m_a+\beta h\right)
+\frac{\left(\beta J\right)^2}{2}\sum_{a\neq b}^{1,n}q_{ab}S_an_aS_bn_b
\nn
\EEA
where the sum in the one-site partition $Z'$ 
is taken over all the possible values 
of the $n$ spins $S_a$ and the
$n$ occupation numbers $n_a$.
Here and everywhere else in this paper we use the abbreviation
\BEQ
\Kt\equiv K+\frac{\beta J^2}{2}
\EEQ
The saddle point equations, coming from the extremization of 
Eq. (\ref{f:62a}),
 give the self-consistent relations
\BEA
&&q_{ab}=\left<S_an_aS_bn_b\right>
\label{f:62d}
\\
&&\rho_a=\left<n_a\right>
\label{f:62e}
\\
&&m_a=\left<S_an_a\right>
\label{f:62f}
\EEA
 where $\left<\ldots\right>$ is the average computed over the measure
$\exp\left(-\beta H'\right)$.

The parameter $\rho_a$ represents the density of the replica $a$.
In the thermodynamic limit this is equal to 
$1/N\sum_in_i={\overline{n_i}}$ which, 
in the Hamiltonian (\ref{f:61a}), is  only coupled to the chemical potential
$\mu$, that is a replica independent quantity.
The density is therefore equal for each replica: $\rho_a=\rho  , \ \ 
 \forall a=1,\ldots,n$.

The same holds for $m_a$ which is coupled to the external field $h$:
$m_a=m , \ \ \ \forall  a= 1,\ldots,n$.
It generally holds that one index quantities are replica invariant. 
\cite{MPV87}

%%%%%%%%%%%%%%%%%%%%%%%%%%%%%%%%%%%%%%%%%%%%%%%%%%%%%%%

		\subsection{Replica Symmetric Solution and 
		  Stability of the Paramagnetic Phase}
		\label{ss:RSsol}

The replica symmetric free energy is obtained by evaluating Eq. (\ref{f:62a})
at $q_{ab}=q_0$, for $a\neq b$, 
$\rho_a=\rho$ and $m_a=m$ for every $a,b$ and it reads
\cite{SNAJPF97} 
\BEQ
\beta f= \frac{\beta \Kt}{2}\rho^2
+\frac{\beta J_0}{2} m^2 -\frac{(\beta J)^2}{4} q_0^2
-\beta J \int_{-\infty}^{\infty}\hspace*{-.3 cm}\d y~P_0(y)~\phi_0(y)
\EEQ
with 
\BEA
\Theta_0&\equiv &\frac{(\beta J)^2}{2}(\rho-q_0) + \beta K \rho + \beta \mu
\\
\nn
&=&\beta\Kt\rho+\beta\mu-\frac{(\beta J)^2 q_0}{2}
\\
P_0(y)&\equiv& \frac{1}{\sqrt{2\pi q_0}}\exp\left[-\frac{
(y-m J_0/J-h/J)^2}{2~q_0}\right]
\\
\phi_0(y)&\equiv& 
\frac{1}{\beta J}\log\left(2+2~e^{\Theta_0}\cosh\beta J y\right).
\EEA

The order parameters $\rho$, $q_0$ and $m$ satisfy the following
 self-consistency equations (see App. \ref{app:BEGC1} for the details
of derivation):
\BEA
&& \rho=\int_{-\infty}^{\infty} \d y~P_0(y)
~\tilde{\rho}(y)
\label{f:621a}
\\
&& q_0=\int_{-\infty}^{\infty}\d y~P_0(y)
~\tilde{m}^2(y)
\label{f:621b}
\\
&&m=\int_{-\infty}^{\infty}\d y~P_0(y)
~\tilde{m}(y)
\label{f:621c}
\EEA
with the following definitions
\BEA
\tilde{\rho}(y)\equiv\frac{\cosh \beta J y}{e^{-\Theta_0}+
\cosh \beta J y}
\label{f:621d}
\\
\tilde{m}(y)\equiv\frac{\sinh \beta J y}{e^{-\Theta_0}+\cosh \beta J y}
\label{f:621e}
\EEA

The eigenvalues of the Hessian of Eq. (\ref{f:62a}) computed in the RS 
approximation
are derived in appendix \ref{app:BEGC1}, where the stability analysis is
 carried out.

In the case $J_0=h=0$ only  the paramagnetic 
solution ($q_0=0$) is stable, so that 
Eqs. (\ref{f:621a})-(\ref{f:621c}) reduce to
\BEQ
q_0=0 \hspace*{.1 cm} ;  \hspace*{ .5 cm}
 \rho=\frac{1}{1+e^{-\beta\Kt\rho-\beta\mu}} 
 \hspace*{ .1 cm} ;  \hspace*{ .5 cm} m=0 \ .
\label{f:621f}
\EEQ
and the  eigenvalues of independent modes 
 are
\BEA
&&\Lambda_0=(\beta J)^2\left[1-(\beta J)^2\rho^2\right]
\label{l2}\\
&&\Lambda_{1}=\beta|\Kt|\left[1-\beta\Kt\left(1-\rho\right)\rho\right]
\label{l1}
\EEA
where $\Lambda_0$ is connected to $\delta q~ \delta q$ fluctuations, 
whereas $\Lambda_1$ is the stability eigenvalue of 
the density-density fluctuations.
In the above formulas we have considered both the case in which $\Kt>0$ and
$\Kt<0$.
The last one  occurring only if the particle-particle interaction $K$ is 
negative  and when $T$ is bigger than $-1/(2K)$. 

The lines $\Lambda_0=0$ and $\Lambda_1=0$ delimit
 the stability region for the RS
 solution on the
 phase diagram.
In the $T$-$\rho$ phase diagram the stable region is for
\BEA
&&T>J\rho
\\
&&T>\frac{K}{2}\rho(1-\rho)\left[1+\sqrt{1
+\frac{J^2}{K^2}\frac{2}{ \rho(1-\rho)}}\right]
%T>\frac{K}{2}\rho(1-\rho)+\sqrt{\frac{[K \rho(1-\rho)]^2}{4}+
%\frac{J^2\rho(1-\rho)}{2}}
\EEA
\noindent
corresponding, respectively, to $\Lambda_0>0$ and  $\Lambda_1>0$.

\begin{figure}[t!]
\begin{center}
{\includegraphics*[width=.49\textwidth]{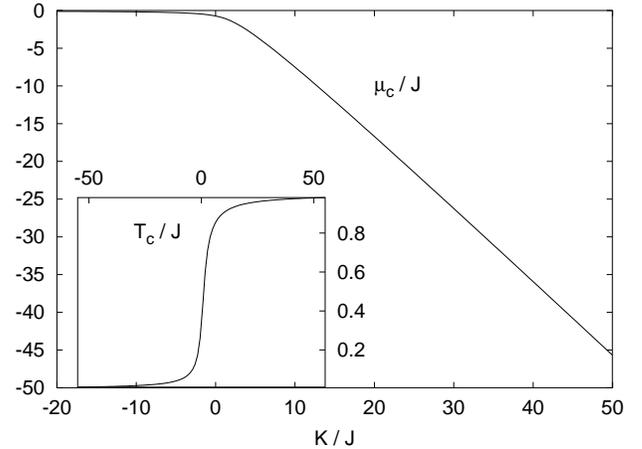}}
\end{center}
\protect\caption{
\small{The values of the thermodynamic parameters (in units of $J$) 
at the tricritical 
point as a function of the particle-particle coupling constant. 
The critical density is
equal to $T_c$. }}
\label{fig:tric}
\end{figure}

For any value of $K$,
there is one intersection between $\Lambda_0=0$ and $\Lambda_1=0$,
namely
the  tricritical point:
\BEA
&&\frac{T_c}{J}=\rho_c
=\frac{J}{2 K}\left(-\frac{3}{2}+\frac{K}{J}
+ \sqrt{\frac{K^2}{J^2}-\frac{K}{J}+\frac{9}{4}}\right)
\label{f:622b}
\\
&&\frac{\mu_c}{J}=  -\frac{1}{2}-\rho_c\left[
        \frac{K}{J}+\log\left(\frac{1}{\rho_c}-1.\right)
        \right]
\label{f:622c}
\EEA
where $\mu_c$ is obtained from the paramagnetic expression 
(\ref{f:621a}-\ref{f:621d})
for $\rho$.
In table \ref{tab1}
we list some values of interest of the tricritical values and in Fig. 
\ref{fig:tric}  we plot their behavior as function  of the particle-particle 
coupling constant. 

 As $K$ decreases, the critical temperature goes to zero. As we will
see in the following, in Sec. \ref{ss:PhDi} where we study the phase
diagrams, this implies that the phase diagram region of phase
coexistence progressively reduces itself.  The critical value of the
chemical potential grows to zero as $K$ tends to $-\infty$: in order
to contrast large particle repulsion a larger chemical potential (in
the limit non-negative) is needed.

 \begin{table}[t!]
\begin{center}
\begin{tabular}{|c|c|c|c|c|c|}
\hline 
 $K/J$ & $\infty$ & 1  & 0  & -1 & $-\infty$
\\
\hline  
 {$T_c/J=\rho_c$} &1 & {1/2} & 1/3 & 0.21922 & 0
\\
\hline  
 $\mu_c/J$ & $-\infty$ & -1 & -0.73105  & -0.55923 & 0
\\
\hline
\end{tabular}
\caption{\small{Critical values of the thermodynamic parameters for
some specific particle-particle interaction values. Parameters are
expressed in units of $J$.}}
\label{tab1}
\end{center}
\end{table}

\subsubsection{Thermodynamic Observables in the Replica Symmetric Solution}

The internal energy and the entropy take the form:

\BEQ
u=\frac{K+\beta J^2}{2}\rho^2 - J_0~ m^2 - h~ m -\mu~\rho 
+\frac{\beta J^2}{4}q_0^2
\EEQ
\BEA
s&=&-\frac{(\beta J)^2}{4}(\rho- q_0)^2-\rho~ \Theta_0
\\
&&+\beta J \int_{-\infty}^{\infty}\d y~P_0(y)~
\left[\phi_0(y)-y~{\tilde{m}}(y)\right]
\nn
\EEA
Integrating by part we can derive:
\BEQ
\int_{-\infty}^{\infty}\d y~P_0(y)~y~{\tilde{m}}(y)=\beta J(\rho- q_0)
\EEQ
The entropy, then, becomes
\BEA
s&=&-\frac{(\beta J)^2}{4}(\rho- q_0)^2 -\rho~ \Theta_0 
\\
\nn
&&-(\beta J)^2
q_0~(\rho - q_0)+\beta J\int_{-\infty}^{\infty}\d y~P_0(y)~\phi_0(y)
\EEA

\subsection{Low Temperature Behavior of the RS Solution}
		\label{ss:RSzeroT}
It can be of some help for later purposes and also to make a comparison
with existing results, to probe how the observables behave in the very low
temperature limit, even in this case for which the SG solution is
unstable. We simplify the discussion to the case $J_0=h=0$. 
Putting $z=y/\sqrt{q_0}$ as integration variable in the above expressions
the entropy can be written as
\BEA
s&=&\frac{(\beta J)^2}{4}(\rho- q_0)^2-\rho~\Theta_0
\\
\nn
&&+
\beta J\int_{-\infty}^{\infty} \frac{dz}{\sqrt{2\pi}}e^{-z^2/2}
\left[\phi_0(z)-\sqrt{q_0}~z~\tilde{m}(z)
\right]
\EEA

We define
\BEA
&& a\equiv\frac{\Theta_0}{\beta J}
\\
&&C\equiv \beta J(\rho -q_0)
\\
&&\Delta I\equiv\int_{-\infty}^{\infty}\hspace*{-2 mm} \frac{dz}{\sqrt{2\pi}}e^{-z^2/2}
\left\{\phi_0(z)-\sqrt{q_0}~z~\tilde{m}(z)
\right\}
\EEA

We consider two cases separately in the zero temperature limit. 

\subsubsection{Full system: $a\geq 0$}
\BEA
&&\lim_{\beta \to \infty}\rho=1
\\
&&\lim_{\beta \to \infty}C=\sqrt{\frac{2}{\pi}}
\\
&&\lim_{\beta \to \infty} a= \frac{1}{\sqrt{2\pi}}+\frac{K+\mu}{J}
\\
&&
 \Delta I =  \Theta_0 +O(T)
\\
&&s = -\frac{1}{2\pi} +O(T)
\EEA

\subsubsection{Partially diluted system: $a<0$}
\BEA
&&\lim_{\beta \to \infty}\rho=\overline{\rho}<1
\\
&&\lim_{\beta \to \infty}C=\overline{C}
\\
&&\lim_{\beta \to \infty} a= \overline{a}\equiv
 \frac{\overline C}{2}+\frac{K\overline{\rho}+\mu}{J}
\EEA
with $\overline{\rho}$ and $\overline{C}$ given by the zero temperature
self-consistency equations:
\BEA
&&\overline{\rho}=\mbox{erfc}
\left(-\frac{\overline a}{2~{\overline \rho}}\right)
\\
&&\overline{C}=\sqrt{\frac{2}{\pi~\overline{\rho}}}
\exp\left(-\frac{\overline a}{\sqrt{2~\overline{\rho}}}\right)
\EEA
Notice that as $a\to 0$ one obtains from the above equations:
$\overline{\rho}=1$ and $\overline{C}=\sqrt{2/\pi}$.

Finally  one gets
\BEA
&& \Delta I =  \overline{\rho}~\Theta_0 + (1-\overline{\rho})\log 2
-\overline{C}~\overline{a}~\log 2 + O(T)
\\
&&s =  -\frac{{\overline{ C}}^2 }{4} + (1-\overline{\rho})\log 2 
-\overline{C}~\overline{a}~\log 2 +O(T)
\EEA

For fixed $K$ and $J$  
there is, thus,  a limiting (negative) value of the chemical potential
below which the
lattice will not be full at zero temperature. For
\BEQ
\frac{\mu^{\star}}{J}<-\frac{1}{\sqrt{2\pi}}-\frac{K}{J}
\EEQ
the density (and the overlap) is equal to $\overline{\rho}<1$.

In any case the asymptotic value of $C$ is finite, yielding
\BEQ
\rho-q_0 \sim T \ .
\EEQ

%%%%%%%%%%%%%%%%%%%%%%%%%%%%%%%%%%%%%%%%%%%%%%%%%%%%%%%
	
	\section{A Variational Method for the Full Replica Symmetry Breaking 
		Solution in Disordered Systems}
    	\label{s:var}
To clarify which kind of transition takes place we improve the study
 of the static properties of the SG phase of the mean-field RBEGC,
 making use of the FRSB Parisi Ansatz.  The first thing to notice is that
 the stable SG phase is always of FRSB type. The transition between
 the PM phase and the SG phase can be either of the SK-type or, below a given
 $T_c$, discontinuous with a jump in the entropy and hence a latent
 heat. Moreover for a certain range of parameters, the two phases, PM and SG,
 coexist. For any parameter choice we find no evidence for a
 $p$-spin-like transition with discontinuous order parameter.

The aim is to evaluate the $n\to 0$ limit in Eq. (\ref{f:62a}) with
 the {\it Ansatz} that the structure of the matrix $Q$ follows a FRSB
 scheme.  In order to be as general as possible, we shall use the RSB
 scheme introduced for the SK model by de Dominicis, Gabay and Orland,
 \cite{DGOJPL81,DGDJPA82} which besides the Edwards-Anderson order
 parameter \cite{EAJPF75} also involves the anomaly to the linear
 response function, otherwise called Sompolinsky's anomaly.
\cite{SPRL81} The more usual Parisi's RSB scheme is, if necessary,
 eventually 
recovered by 
a proper gauge fixing, once that the limit of infinite number of replica 
symmetry breakings has been taken  [see below Eq. (\ref{gaugeP})]. 

By applying the RSB scheme infinite times and introducing the two 
functions
$q(x)$ ({\em overlap} function) and $\Delta(x)$
({\em anomaly} function), $0\leq x\leq 1$, 
 the free energy functional, 
Eq. (\ref{f:62a}), becomes:

\BEA
&&\beta f=\rho^2 \frac{\beta \Kt}{2}
+\frac{\beta}{2}J_0~m^2-\frac{(\beta J)^2}{4}q(1)^2
\label{f:63a}
\\
&&\ \ -\frac{\beta J}{2} \int_0^1\ dx \ q(x)~\dot\Delta(x)
-\beta J\int^{\infty}_{-\infty}\dy~ P_0(y) ~\phi(0,y) \ ,
\nn
\EEA
\noindent
where $P_0(y)$ is defined as
\BEQ
P_0(y)\equiv \frac{1}{\sqrt{2\pi q(0)}}
\exp\left\{-\frac{\left(y-(h+J_0~m)/J\right)^2}{2 q(0)}\right\}
\label{f:P0}
\EEQ
\noindent and  $\phi(0,y)$ is the solution, evaluated at $x=0$, of the Parisi 
parabolic equation~\cite{PJPA80}
\BEQ
\dot\phi(x,y)=-\frac{\dot q(x)}{2}\phi''(x,y)+
\frac{\dot\Delta(x)}{2}\phi'(x,y)^2 \ ,
\label{f:63b}
\EEQ
\noindent
with the boundary condition at $x=1$
\begin{equation}
\phi(1,y)=\phi_1(y)\equiv(\beta J)^{-1}
\log\left(2+2 e^{\Theta_1}\cosh\beta J y\right) \ ,
\label{f:63c} 
\end{equation}
\noindent
and
\BEA
&&\Theta_1\equiv \frac{(\beta J)^2}{2} [\rho-q(1)] +\beta(\mu+K \rho)
\label{f:63d}
\\
\nn
&&\hspace*{1 cm}=\beta\Kt\rho+\beta\mu-q(1)\frac{(\beta J)^2}{2} \ .
\EEA

The  overlap $q(x)$, the density of occupied
sites $\rho$  and  the  anomaly
 $\Delta(x)$ are the order parameters.
\footnote{In the dynamical approach of Sompolinsky~\cite{SPRL81,SZPRB82}
 to the spin glass problem $\Delta$ represents, in the
Fluctuation-Dissipation Relation (FDR), the anomaly in the linear
response with respect to the equilibrium value}  We have used the
standard notation and denoted derivatives with respect to $x$ by a dot
and derivatives with respect to $y$ by a prime.  In this notation
Sompolinsky's $\Delta'$ in Ref. \onlinecite{SPRL81}  becomes our $T\dot\Delta$.

The Parisi  equation (\ref{f:63b}) can be included into a
 free energy variational functional via the 
Lagrange multiplier $P(x,y)$ and the initial condition at $x=1$, 
Eq.  (\ref{f:63c}),
via $P(1,y)$. \cite{SDJPC84} The free energy functional is then
\BEA
\beta f_{\rm v}&=&\beta f
\\
&+&\beta J\int_{-\infty}^{\infty}\dy P(1,y)\left[
\phi(1,y)-\phi_1(y)\right]
\nn
\\
&-&\beta J\int_0^1 \dx \int_{-\infty}^{\infty}\dy P(x,y)
\Bigl[\dot\phi(x,y) 
\nn
\\
&&\hspace*{2 cm}+
\frac{\dot q(x)}{2}\phi''(x,y)- \frac{\dot\Delta(x)}{2}\phi'(x,y)^2 \Bigr]
\ .
\nn
\EEA
with $P_0(y)$  and $\phi_1(y)$ defined in Eqs. ({\ref{f:P0}, \ref{f:63c}).

By such a construction $f_{\rm v}$ is stationary with respect to  variations of
$P(x,y)$, $P(1,y)$, $\phi(x,y)$, $\phi(0,y)$, 
$q(x)$, 
$\dot\Delta_q(x)$ and deriving with respect to $\rho$.
Variations of $P(x,y)$ and $P(1,y)$ simply give back 
Eqs. (\ref{f:63b}) and (\ref{f:63c}).
Stationarity with respect to variations of $\phi(x,y)$ and $\phi(0,y)$
leads to a partial differential equation for $P(x,y)$:
\BEQ
\dot P(x,y)=\frac{\dot q(x)}{2}P''(x,y)
+\dot\Delta(x)\left[P(x,y)\phi'(x,y)\right]' \ ,
\label{f:63f}
\EEQ
and to the boundary condition at $x=0$
\BEQ
P(0,y)=P_0(y)\ .
\label{f:63g}
\EEQ

Eventually, variations of $f_{\rm v}$ with respect to
 $q(x)$, $\dot\Delta(x)$ and the derivative  with respect to $\rho$
 lead to
\BEA
&&\hspace*{-5 mm}\Delta(x)=-\beta J \left[\rho-q(1)\right] + \hspace*{-.2 cm}
\int_{-\infty}^{\infty}\hspace*{-.4 cm} \dy P(x,y)~\phi''(x,y)
\label{f:63h}
\\
&&\hspace*{-5 mm}q(x)=\int_{-\infty}^{\infty} \dy P(x,y)\ \phi'(x,y)^2
\label{f:63i}
\\
&&\hspace*{-5 mm}\rho= \int_{-\infty}^{\infty} dy\ P(1,y)\ 
\frac{\cosh \beta J y}{e^{-\Theta_1}+\cosh \beta J y}
\label{f:63l}
\EEA
with  $\Delta(1)=0$, the anomaly at the shortest time-scale, 
corresponding to $x=1$, being zero by construction: the Fluctuation-
Dissipation Theorem (FDT) holds at short time-scales.

The Lagrange multiplier $P(x,y)$ represents the distribution of local
fields. One may indeed associate a given overlap $q(x)$ with a time
scale $\tau_x$ such that for times of order $\tau_x$ states with an
overlap equal to $q(x)$ or greater can be reached by the system (these
time-scales completely decouple in the thermodynamic limit).  In this
picture the $P(x,y)$ becomes the probability distribution of frozen
local fields $y$ at the time scale labeled by $x$. \cite{SDJPC84}

For the numerical treatment of the FRSB equations and also to allow
for a clearer physical interpretation of the functions that we are
analyzing, we define the {\em local magnetization}
$m(x,y)\equiv\phi'(x,y)$, whose differential antiparabolic equation we
derive from Eqs. (\ref{f:63b}), (\ref{f:63c}) as
\BEA
&&\hspace*{-5mm}
\dot m(x,y)=-\frac{\dot q}{2} m''(x,y)+\dot\Delta(x)~m(x,y)~m'(x,y)  \ ,
\label{f:63m}
\\
&&\hspace*{-5mm}
m(1,y)=m_1(y)=\frac{ \sinh(\beta y)}{e^{-\Theta_1}+\cosh(\beta y)} \ .
\label{f:63n}
\EEA

The average equilibrium
magnetization $m$ can, then,  be computed in terms of the local magnetization
at $x=0$:
\BEQ
m=-\frac{\p f}{\p h}=\int_{-\infty}^{\infty}\dy P(0,y)~ m(0,y) \ .
\label{f:63n2}
\EEQ

Another useful relation for the numerical evaluation of the order parameters
and the thermodynamic  functions built on them is got from the computation
of the term $\int_0^1\dx\int_{-\infty}^{\infty}\dy P(x,y)\dot\phi(x,y)$ in two 
different
ways:
once using Eq. (\ref{f:63c}) and, in the other case,
 applying Eq. (\ref{f:63f}) and integrating by part.
From the comparison of the two results  the following equation is 
obtained:
\BEA
&&\hspace*{-5mm}\int_{-\infty}^{\infty}\dy~ P(1,y) ~\phi(1,y)-
\int_{-\infty}^{\infty}\dy ~P(0,y)~ \phi(0,y) \ \ 
\label{f:63o}
\\
\nn
&&\hspace*{4 cm}=-\frac{1}{2}\int_0^1 \dx~q(x)~\dot\Delta(x)
\EEA

Deriving Eq. (\ref{f:63h}) with respect to $x$ yields 
\BEQ
\int_{-\infty}^{\infty}\dy~ P(x,y)~ m'(x,y)^2 =1 
\label{f:63p}
\EEQ
valid for every $x$ and guaranteeing the marginal stability of the FRSB Ansatz.

The coupled equations (\ref{f:63f}), (\ref{f:63m}), with border conditions
(\ref{f:63g}), (\ref{f:63n}) are the FRSB equations.
In the following we are going to show how they can be numerically
solved
and what are the gauges we will use in order
 to get the most general
SG phase and the order parameters characterizing the transition to it.

Differentiating once more Eq. (\ref{f:63p}) one finds 
\BEQ
-\frac{\dot \Delta}{\dot q}=
\frac{\int _{-\infty}^{\infty}\dy~ P(x,y)~ m''(x,y)^3}
{\int_{-\infty}^{\infty}\dy~ P(x,y) ~m'''(x,y)^2}
\label{f:csi}
\EEQ
that will become useful is the following when we will discuss the 
choice of the gauge to perform the numerical 
computation (see Sec. \ref{ss:Gauge}).
%%%%%%%%%%%%%%%%%%%%%%%%%%%%%%%%%%%%%%%%%%%%%%%%%%%%%%%

 		\subsection{Numerical Integration of the FRSB Equations:
		the Pseudo-Spectral method}
 		\label{ss:PSint}
	
In order to study the low temperature regime of the RBEGC in the 
limit of a large number of clauses we have numerically integrated the 
FRSB equations (\ref{f:63f})-(\ref{f:63m})
to determine $q(x)$, $P(x,y)$ and $m(x,y)$.
We followed the iterative scheme of Refs. [\onlinecite{SDJPC84, NJPC87}], 
but with an improved numerical method which allows for very
accurate results for all temperatures (see Refs. 
[\onlinecite{CLPJPA02, CRPRE02}]). 

We start from an initial guess for $q(x)$, $\Delta(x)$ and $\rho$
then $m(x,y)$, $P(x,y)$ and the
associated $q(x)$ are computed in the order as:
\begin{enumerate}
\item 
        Compute $m(x,y)$ integrating from $x=1$ to $x=0$ Eqs. 
        (\ref{f:63m}) with initial condition (\ref{f:63n}).
\item
        Compute $P(x,y)$ integrating from $x=0$ to $x=1$ Eqs. 
        (\ref{f:63f}) with initial condition (\ref{f:63g}).
\item
        Compute $q(x)$, $\Delta(x)$ and $\rho$ 
	using Eqs. (\ref{f:63i}-\ref{f:63l}).
\end{enumerate}

The steps $1\,\to\, 2\,\to\, 3$ are repeated until a 
reasonable convergence
is reached, typically we require a
mean square error on $q$, $P$ and $m$  of the 
order $O(10^{-6})$ and we checked that the identities (\ref{f:63o}),
(\ref{f:64c}) and $1-T=\overline{q}$ were satisfied to this precision as well.
The number of iterations necessary are a few hundreds. 
The core of the integration scheme is the integration of the partial 
differential equations (\ref{f:63f}) and (\ref{f:63m}). 
In previous works this was carried out through 
direct integration in the real space which requires a large 
grid mesh to obtain precise results. 
To overcome this problem we move to the Fourier space where we can apply
a pseudo-spectral~\cite{Orszag71} method of integration.

Indicating by ${\bf{FT}}[o](x,k)$ the Fourier transform
of function $o(x,y)$,
\BEQ
{\bf{FT}}[o](x,k)=\frac{1}{N_y}\int_{-N_y/2}^{N_y/2}dy~e^{-iky}~o(x,y)    \ ,
\label{f:731}
\EEQ
the FRSB Eq. (\ref{f:63c}), written in terms of the wave number $k$,
becomes
\BEQ
\dot m(x,k)=k^2\frac{\dot q(x)}{2}~ m(x,k)+
ik~\frac{\dot \Delta(x)}{2}{\bf FT}[m^2](x,k)
\label{f:731a}
\EEQ
and the FRSB Eq. (\ref{f:63f}) takes the form
\BEQ
\dot P(x,k)=-k^2\frac{\dot q(x)}{2}~P(x,k)
+ik~\dot \Delta(x)~{\bf FT}[P~m](x,k)
\label{f:731b}
\EEQ
For each value of 
 $k$ these are ordinary differential equations
 which can be integrated using standard methods.
To avoid the time consuming calculation of the convolution in the nonlinear
terms we use the  pseudo-spectral code on a grid mesh of $N_x\times N_y$
points, covering the $x$ interval $[0,1]$ and the $y$ interval 
$[-y_{\rm max},y_{\rm max}]$.
The truncation of the wave number may bring to anisotropic effects for large 
$k$.  De-aliasing  is, thus, performed by 
a $N_y/2$ truncation,
 which ensures a better isotropy of the numerical treatment.
Eventually, the $x$ integration is carried out by means of a third
 order Adam-Bashford scheme which reduces the number of fast Fourier 
transforms (FFT) calls.
\cite{Ferziger96}

 Typical values used are $N_x=500 \div 1000$, $N_y = 1024 \div 4096$
and $y_{\rm max} = 24 \div 48$.  The difference between the values for
$N_x$ and the values for $N_y$ comes from the fact that, if the
solution in $y$ is smooth enough, only a few low wave numbers $k$ are
exited.  The value of the parameter $y_{\rm max}$ fixes the $y$ range
where the solution is assumed different from zero.  Indeed, in the
numerical algorithm is assumed that $P(x,y)=m(x,y)=0$ for $|y|>y_{\rm
max}$.  This explains the rather large value of $y_{\rm max}$ used.

		\subsection{Choice of Gauge}
		\label{ss:Gauge}

The  solution to the spin-glass mean-field models obtained using the scheme
of Sommers is overconstrained and the 
functional expression of $\dot\Delta(x)$ can be, thus, 
chosen in different ways,
 selecting, in this way,  a gauge for the order parameters.
The usually studied gauges are
\BEA
&&\dot\Delta(x)=-\beta J~ x~\dot q(x) \hspace*{ 1 cm} {\mbox{Parisi gauge
\cite{PJPA80}}}
\label{gaugeP}
\\
&&\dot\Delta(x)=-\Delta(0) \hspace*{ 1.8 cm} {\mbox{Sommers gauge
\cite{SDJPC84}}}
\label{gaugeS}
\EEA
\noindent where the anomaly is given by the stationarity 
Eq. (\ref{f:63h}) at $x=0$. We rewrite it here in terms of the local 
magnetization
\BEQ
\Delta(0)=-\beta J \left[\rho-q(1)\right]
 + \int_{-\infty}^{\infty}\hspace*{-.2 cm} \dy~ P(0,y)~ m'(0,y)
\label{f:D0g}
\EEQ
The anomaly at the largest time-scale is gauge invariant since it
depends on the Edwards-Anderson parameter and on the derivative of the
equilibrium local magnetization. $\Delta(x)$ measures the violation of
the linear response on the time-scale labeled by $x$.  As such, it's a
decreasing function of $x$, being zero at the shortest time-scale
($x=1$) where the dynamical relaxation of the system has not yet
fallen out of equilibrium, and maximum at the longest time-scale
$x=0$.

Other ``intermediate'' gauges can be introduced for numerical purposes, e.g.
\BEQ
\dot\Delta(x)= - \gamma \Delta(0)(1-x)^{\gamma-1}, \ \ \ \gamma=1, 2, \dots
\label{gaugeI}
\EEQ
that, for $\gamma =1$, is the Sommers gauge.

In practice, the most common choice is the Parisi gauge
$\dot\Delta(x)=-\beta J~ x~\dot q(x)$, useful from a numerical point
of view, since for $x$ larger than a certain critical value $x_c$, at
which the overlap function displays a cusp and reaches the plateau
value $q_{EA}$, $\dot q(x)$ tends to zero and the integration domain
can, thus, be reduced.  For $T\to 0$, however, the overlap function in
this gauge tends to a step function, yielding, for some $x$, a
diverging factor $\dot q(x)$ in both terms of the rhs of equations
(\ref{f:731a}) and (\ref{f:731b}).

For this reason the alternative Sommers gauge can be adopted
$\dot\Delta(x)=-\Delta(0)$, so that the non linear terms of the rhs of
Eqs. (\ref{f:731a}) and (\ref{f:731b}) is automatically kept finite.
The function $q(x)$ comes out to be a smoother function of $x$ in this
gauge, not only continuous and monotonous,
but also without any cusp at any $x$.  When
the external magnetic field is zero $\dot q$ is still divergent, but
only exactly at $x=0$, thus yielding no sharp change in convexity,
whereas for $h\neq 0$ even this divergence disappears and the FRSB
equations do not display any integration problem because of $q(x)$
behavior.

\subsubsection{Gauge dependent parameters and physical observables}

The Parisi gauge is, thus, advantageous down to temperatures where it
starts bringing numerical instabilities.  For lower temperatures the
Sommers gauge (or some similar, e.g. Eq. (\ref{gaugeI})) stabilizes
the results.  To choose a gauge means to select the $x$ behavior of the
order parameter functions will be different. This will not change,
anyway, the physical quantities, such as density, Edwards-Anderson
parameter ($q_{\rm EA}=q(1)$), energy, entropy, etc., that we are
going to show in the next section,

Also the overlap probability distribution $P(q)$ is gauge invariant.
However, one has to consider the fact that the definition $P(q)=d
x(q)/d q$ only holds in the Parisi gauge.  To build a general
definition of $P(q)$ as the derivative of a cumulative distribution we
can use Eq. (\ref{f:csi}) and introduce the cumulative function $\xi $
\BEQ \xi(x)\equiv -\frac{T\dot\Delta(x)}{\dot q(x)}
\label{s:csi2}
\EEQ
where $x$ here is the gauge-dependent RSB parameter and $\Delta$ and
$q$ are the gauge-dependent order parameter.  In the Parisi gauge is
$\xi(x)=x$, and we get back the usual definition.  In the Sommers
gauge is $\xi(x)=T\Delta(0)/\dot q(x)$ and $\xi(q)$ can be obtained,
e.g. parametrically, using Eq. (\ref{f:63i}).
Thus, the  expression of the overlap distribution has to be derived
as 
\BEQ
P(q) = \frac{d \xi(q)}{d q}
\EEQ

\subsubsection{Parisi gauge as fluctuation-dissipation ratio}

Following a dynamical interpretation of the order parameters $q(x)$
and $\Delta(x)$, respectively as spin-spin correlation function and
anomaly in the susceptibility at time-scale $x$, the Parisi gauge is
actually a rewriting of the Fluctuation-Dissipation Relation.  That is
the generalization of the FDT for aging
systems dynamically stuck out of equilibrium, such as spin-glasses,
where the coefficient between correlation and response functions is
not $T$ but some function of the time-scale on which the system is
relaxing (often referred to as {\em effective temperature}).  
We saw in Eq. (\ref{f:chi_x}) that $\Delta(x) = \chi(x) - \chi(1) $ is
the anomaly of the susceptibility with respect to the linear response
value (i.e. the ZFC susceptibility). Thus, $\dot\Delta(x)$ is the
anomalous response function, whereas $\dot q$ is the derivative of the
correlation function and $T_{\rm eff}=1/(\beta x)$ the effective
temperature. The time dependence of $T_{\rm eff}$ is expressed through
the time-scale index $x$.

%%%%%%%%%%%%%%%%%%%%%%%%%%%%%%%%%%%%%%%%%%%%%%%%%%%%%%% 	
	\section{Thermodynamic Observables}
     	\label{ss:BEGCTD}

All thermodynamic quantities can be written in terms of 
the order parameter functions derived by solving
Eqs. (\ref{f:63c}), (\ref{f:63f}), namely  overlap, density, anomaly,
local magnetization and distribution of local fields.
\subsubsection{Internal energy}
The internal energy $u=\p \beta f/\p\beta$
can be computed either taking the derivative of Eq. (\ref{f:62a}) 
(and then breaking
the replica symmetry infinite times) or directly deriving 
 Eq. (\ref{f:63a}).

In the first case the energy comes out to be  
\BEA
u&=&\lim_{N_B\to\infty}
\lim_{n\to 0}\left[-\frac{K+\beta J^2}{2}\frac{1}{n}\sum_a \rho_a^2
-J_0\frac{1}{n}\sum_a m_a^2\right.
\nn
\\
\nn
&&\left.\hspace*{1cm}-h\frac{1}{n}\sum_a m_a
-\mu\frac{1}{n}\sum_a \rho_a
-\frac{\beta J^2}{2}\frac{1}{n}\sum_{a\neq b}~q_{ab}^2
\right]
\nn
\\
&=&- \frac{K+\beta J^2}{2} \rho^2
-\frac{J_0}{2} \ m^2-h \ m -\mu\  \rho \ \ 
\label{f:64a}
\\
\nn
&&\hspace*{2cm}+\frac{\beta J^2}{2} q(1)^2+J \int_0^1 dx~q(x)~\dot\Delta(x)
\EEA
where $N_B$ is the number of replica symmetry breakings.
Otherwise, deriving Eq. (\ref{f:63a}) and using 
 Eqs. (\ref{f:63c},\ref{f:63o}), one finds
\BEA
u&=&-\frac{K+\beta J^2}{2} \rho^2-\mu \rho + \frac{J_0}{2} m^2 
\label{f:64b}
\\
\nn
&&+\beta J^2 q(1)\ \rho 
-\frac{\beta J^2}{2}q(1)^2
-J\int_0^1 \dx~ q(x)~\dot\Delta(x)
\nn
\\
\nn
&&-J\int_{-\infty}^{\infty}\dy P(1,y)~y~m(1,y)
\EEA

The comparison between Eq. (\ref{f:64a}) and Eq. (\ref{f:64b})
yields the relation
\BEA
&&2 \int_0^1\dx ~q(x)~\dot\Delta(x)=\beta J~q(1)~\left[\rho-q(1)\right]
\label{f:64c}
\\
\nn
&&+\frac{J_0}{J}m^2+\frac{h}{J}m
-\int_{-\infty}^{\infty}\dy P(1,y)~y~m(1,y)
\EEA

Taking the limit $\rho\to 1$, with $K=\mu=0$ and choosing the gauge
$\dot\Delta=-\beta J~x ~\dot q(x)$ (corresponding to the Parisi scheme
for RSB) we get the SK formulas \cite{SKPRL75}.

\subsubsection{Entropy density}
The entropy density $s=\beta^2\partial f/\partial \beta$
 can be expressed either
as
\BEA
s&=&-\rho~\Theta_1
-\frac{(\beta J)^2}{4}\left[\rho-q(1)\right]^2
\label{entro}
\\
&&+\beta J\int_{-\infty}^\infty dy~P(1,y)\left[\phi(1,y)-y~m(1,y)\right]
\nn
\EEA
\noindent or, exploiting  Eq. (\ref{f:64c}), as 
\BEA
s&=&-\frac{(\beta J)^2}{4}\left[\rho-q(1)\right]^2-\rho~\Theta_1
-(\beta J)^2~q(1)~\left[\rho - q(1)\right] \label{entro2}
\nn
\\
\nn
&&
-\beta J_0~m^2-\beta h~m+~2\beta J\int_0^1 dx~q(x)~\dot\Delta(x)
\\
&&
+\beta J\int_{-\infty}^\infty dy~P(1,y)~\phi(1,y)
\EEA

\subsubsection{Compressibility}
\begin{figure}[t!]
\begin{center}
{
 \includegraphics*[width=0.49\textwidth]{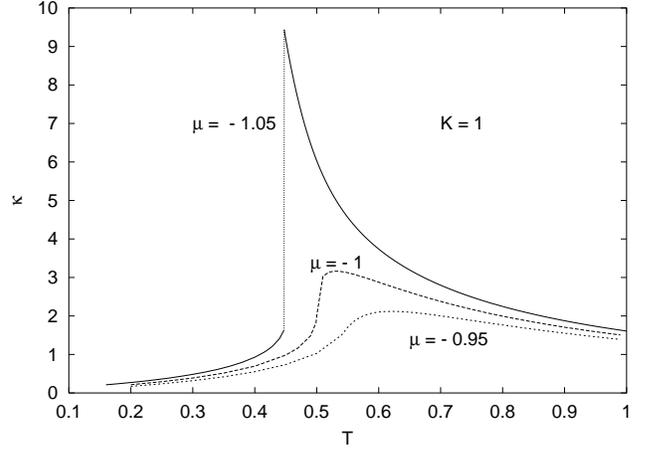}
}
\end{center}
\label{fig:kappa}
\protect\caption{\small{ Compressibility versus temperature for
 $K/J=1$, in units of $J$. The behavior at $\mu$ above $\mu_c/J=-1$ is
 smooth at the transition (right dotted line).  At $\mu_c$ a
 discontinuity starts to develop (middle dashed curve) leading to a
 cusp for $\mu<\mu_c$ (left solid curve). } }
\end{figure}

The compressibility in terms of the density of occupied sites and its
conjugated field, the chemical potential, can be expressed in the FRSB
formulation as:
\BEQ
\kappa = \frac{1}{\rho^2}\frac{\partial \rho}{\partial \mu}
=\frac{\beta}{\rho^2}\left[\rho-\int_{-\infty}^\infty dy~P(1,y)~
\rho^2_1(y)\right] \ ,
\EEQ 
where we define 
\BEQ
\rho_1(y)\equiv\frac{\cosh(\beta J y)}{e^{-\Theta_1}+\cosh(\beta J y)} \ .
\EEQ
In Fig. \ref{fig:kappa} we show the compressibility behavior versus 
temperature
of the system with particle-particle interaction constant $K=J$ for three 
values of the chemical potential.
Respectively above, at and below the critical value $\mu_c$ below which the
 transition happens to be
first order in the Ehrenfest sense.
For $\mu<\mu_c$ a cusp shows up and, decreasing the temperature,  $\kappa$ 
shrinks
to a much lower value as the transition point is crossed.

\subsubsection{Susceptibility}
\begin{figure}[!t]
\begin{center}
{
 \includegraphics*[width=0.49\textwidth]{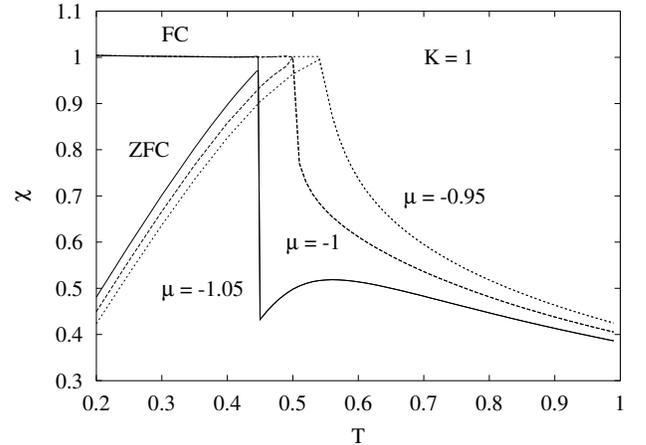}
}
\end{center}
\protect\caption{\small{ Field cooled and zero-field cooled susceptibility for
$K=1$. Above $\mu_c=-1$ FC ($\chi_0$) and ZFC ($\chi_1$) susceptibilities
continuously 
reach the same value at the second order phase transition 
($\chi_0=\chi_1=1$ in absence of external magnetic field), 
whereas for $\mu<\mu_c$}
both a discontinuity takes place and the value of $\chi_1$ never reaches the 
one of $\chi_0$ at the, now first order,
phase transition.} 
\end{figure}      

The  magnetic susceptibility of the RBEGC model at equilibrium  can be 
written either as
\BEQ
\chi_0=\beta J\left[\rho-\int_0^1 dx~q(x)\right]
\label{f:64g}
\EEQ
or, with the help of Eq. (\ref{f:63n2}), as
\BEQ
\chi_0=\frac{\p m}{\p h}=\int_{-\infty}^{\infty}\dy P(0,y)~ m'(0,y) \ ,
\EEQ
whereas the  susceptibility obtained when the system is constrained
to stay in a single minimum of the free energy function comes out to be
\BEQ
\chi_1=\beta J\left[\rho-q(1)\right] \ .
\label{f:64h}
\EEQ

The equilibrium  susceptibility is a function of the average 
${\overline q}=\int_0^1 dx~q(x)$
of the overlap  over all possible values it can take at all time scales
($0\leq x\leq 1$). It corresponds to the Field-Cooled susceptibility.
The second susceptibility, instead, only depends on the Edwards-Anderson
parameter $q_{\rm EA}=q(1)$. It physically expresses the
self-overlap of configurations belonging to the same state. Indeed,
at time scale $\tau_1$, i.e., the shortest time scale,
the system has not yet visited but one metastable state, thus the
response  to a field perturbation only depends  on the self-overlap.
The experimental analogue of $\chi_1$  is the Zero Field Cooled susceptibility.
As a matter of fact, also in that case the system remains in one 
single  state during the cooling down to the SG phase, since it 
is not driven by any external field.

From a dynamical point of view,
the equilibrium 
 susceptibility $\chi_0$
can otherwise be expressed in terms of the function
$\Delta(x)$ defined by Sompolinsky to encode the anomalous response  
to a field
perturbation at large time scales ($x<1$, $\tau_x>\tau_1$
 in the parametric representation used so far). 
The anomaly function is a
 direct way
to measure ergodicity breaking occurring in spin glasses (see Ref.
\onlinecite{MPV87,Fischer91}), even at infinite time ($x=0$).
Defining the susceptibility function at the time-scale labeled by $x$ as
\BEQ
\chi(x)\equiv \beta J[\rho-q(1)] +\Delta(x) \ ,
\label{f:chi_x}
\EEQ
where  $\Delta(x)$ is given by the  stationarity Eq. (\ref{f:63h}),
the equilibrium susceptibility, Eq.  (\ref{f:64g}), can be rewritten as
\BEQ
\chi_0=\chi(0)=\beta J[\rho-q(1)]+\Delta(0) \ .
\EEQ

The anomaly $\Delta(0)$ can, thus,  be interpreted as the difference 
between the  theoretical 
descriptions of the zero-field-cooled and the field-cooled susceptibility.
\BEA
&&\hspace*{-5 mm}\Delta(0)=\chi(1)-\chi(0)=
\beta \left[q(1)-\int_0^1dx~q(x)\right]
\\
\nn
&&\hspace*{3 cm}=\beta(q_{EA}\ \ -\ \ {\overline{q}})\ .
\EEA

\begin{figure}[t!]
\begin{center}
{
\includegraphics*[width=0.49\textwidth]{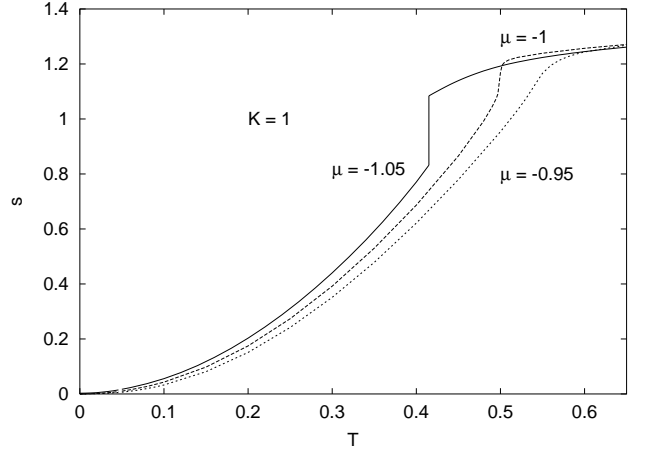}
}
\end{center}
\protect\caption{\small{
Entropy density as a function of temperature for $K=J$.
        For $\mu<\mu_c=-J$
        the entropy is discontinuous at the transition temperature:
	a latent heat is produced/employed at the transition.
	}        }
\label{fig:ent.K1.0}
\end{figure}
\begin{figure}[th!]
\begin{center}
{
\includegraphics*[width=0.49\textwidth]{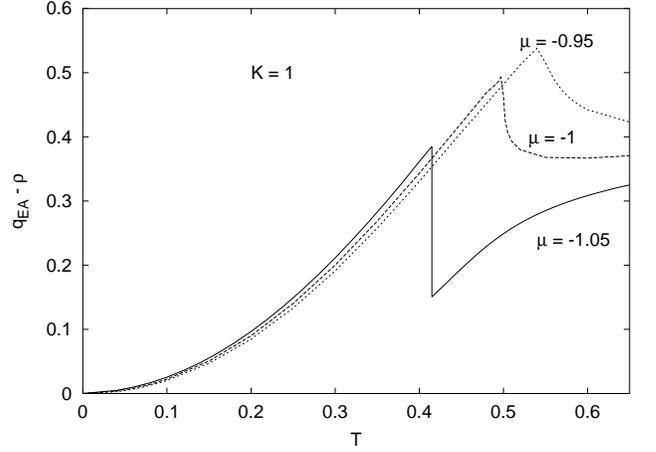}
}
\end{center}
\protect\caption{\small{
$\rho - q(1)$ as a function of temperature for $K=J$.
        For $\mu<\mu_c=-J$
        there is a discontinuity  at the transition temperature.
	}        }
\label{fig:rho_q1.K1.0}
\end{figure}
\begin{figure}[th!]
\begin{center}
{
\includegraphics*[width=0.49\textwidth,height=.40\textwidth]{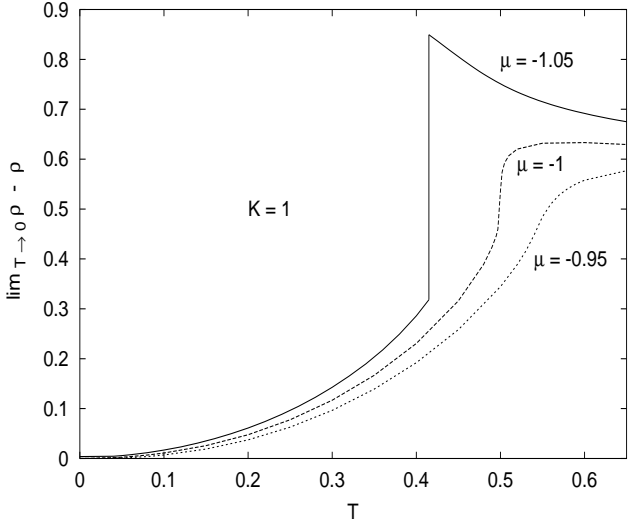}
}
\end{center}
\protect\caption{\small{
${\overline \rho} - \rho $ as a function of temperature for $K=J$.
        For $\mu<\mu_c=-J$
	${\overline \rho}<1$ and 
         there is a discontinuity  at the transition temperature.
	}        }
\label{fig:1_rho.K1.0}
\end{figure}

\subsubsection{ Free Energy Integral 
$\int_{-\infty}^{\infty} \dy~ P(0,y)~ \phi(0,y)$}

The often occurring integral $\int_{-\infty}^{\infty} \dy~ P(0,y)~
\phi(0,y)$ [see Eqs. (\ref{f:63a}) and (\ref{f:63o})] can be expressed
as a function of the local magnetization $m(0,y)$, numerically found
as solution of coupled Eqs.  (\ref{f:63f}), (\ref{f:63m}) $x=0$.  From
the magnetization definition we write the identity \BEQ
\phi(0,y)=\phi(0,0)+\int_0^y dy'~m(0,y') \ , \EEQ
\noindent
that is valid for every $y$.
Combining this with the identity 
$\phi(1,y)-\phi(0,y)=\int_0^1 dx~\dot\phi(x,y)$
and taking the limit $|y|\gg 1$, we get for the constant $\phi(0,0)$ 
the value
\BEA
&&\phi(0,0)=\lim_{|y|\to \infty}\left\{
 T~\Theta_1+|y|-\frac{1}{2}\int_0^1 dx~\dot\Delta(x)\right.
\\
\nn
&&\hspace*{4 cm}\left.-\int_0^y dy'~m(0,y')\right\}
\\
\nn
&&=
T~\Theta_1-\frac{1}{2}\int_0^1\hspace*{-2mm}dx~\dot\Delta(x)
+\int_0^\infty\hspace*{-2mm}dy~
\left[1-m(0,y)\right] \ .
\EEA
This leads to the more convenient expression, for a numerical computation,
\BEA
&&\int_{-\infty}^{\infty} \dy~ P(0,y)~ \phi(0,y)
\\
\nn
&&\hspace*{1 cm}=T~\Theta_1-\frac{1}{2}\int_0^1\hspace*{-2mm} dx~\dot\Delta(x)
+\int_0^\infty\hspace*{-2mm}dy~
\left[1-m(0,y)\right]
\\
\nn
&&\hspace*{3 cm}+\int_{-\infty}^{\infty}\hspace*{-2mm} \dy~ P(0,y)
\int_0^y\hspace*{-2mm} dy'~m(0,y') \ .
\EEA

%%%%%%%%%%%%%%%%%%%%%%%%%%%%%%%%%%%%%%%%%%%%%%%%%%%%%%%

\subsection{Low Temperature Behavior in the FRSB SG Phase}
		
		\label{ss:FRSBLowT}

From the self-consistency Eqs. (\ref{f:63i}-\ref{f:63l}) 
we can get the $T=0$ limits of $q(1)$ and $\rho$.
As in the RS case previously reported (see Sec. \ref{ss:RSzeroT}), 
also for the stable solution
that limit crucially depends  from the sign  of $\Theta_1$ at zero temperature.
Once again we define $a\equiv \Theta_1(T)/(\beta J)$ and 
we sketch the results in the two qualitatively different cases:

\subsubsection{Full lattice at zero temperature: $a>0$}

For $T\to 0$ the last term of the entropy in  Eq. (\ref{entro}) goes as
\BEA
&&\beta J\int_{-\infty}^{\infty}
dy~P(1,y)\left[\phi(1,y)-y~m(1,y)\right]
\label{T0l}
\\
&&\hspace*{2 cm}= \Theta_1(0)+ O(e^-{1/T})
\nn
\EEA
where the boundary functions $\phi(1,y)$ and $m(1,y)$
are given by Eqs. (\ref{f:63c}, \ref{f:63n}).
The Edwards-Anderson parameter and the density both tend to one.
 
At zero temperature there is one single ground state
and  the entropy is, therefore,  zero.
So that,
from Eq. (\ref{entro}), we get
\BEQ
\hspace*{.5cm}0=\lim_{\beta\to\infty}\left\{-\frac{(\beta J)^2}{4}
\left[\rho-q(1)\right]^2+(1-\rho)\Theta_1
\right\}
\label{f:zeros}
\EEQ

Using Eq. (\ref{T0l}) and 
equating the  entropy expression at zero temperature to zero we see that 
 $q(1)$ and $\rho$ go to their $T=0$ value in such a way that
$\rho-q(1)\sim T^s$ and $1-\rho\sim T^{s'}$,with  $s,s'\geq 1$

To analytically determine the value of $s$ and $s'$ is turns out to be
quite complicated.  However, from our numerical computation, we find
that, at low temperatures both the entropy and $\rho-q(1)$ reach the
zero value as $T^2$, thus $s=2$.  In order to satisfy formulas
(\ref{entro}) and (\ref{T01}), then, $1-\rho$ has to decrease as
$T^3$ ($s'=3$), also consistent with our numerical data.

\subsubsection{Partially empty lattice: $a<0$.}

The parameter $\Theta_1$ tends to $-\infty$ at zero temperature and
 $q(1)$ and $\rho$ are less than one,
namely
\BEA
&&\lim_{\beta\to \infty}q(1)=\lim_{\beta\to \infty}\rho\equiv
\overline{\rho}
=2\int_{-\overline{\rho} K/J-\mu/J}^\infty\hspace*{-1.5 cm} dy~P(1,y)
\\
\nn
&&=2\int_{0}^\infty dy~P(1,y-\overline{\rho}K/J-\mu/J)
\EEA

If  $\Theta_1 < 0$, in general,
$e^{\Theta_1}\cosh\beta J y$ will not be much larger than one 
 for any value of $y$
in the argument 
of $\phi(1,y)$ and in $y~m(1,y)$.
The zero $T$ limit  depends on the sign of 
$\alpha_{|y|}\equiv |y|+ a/J$:
\BEQ
\lim_{\beta \to \infty}\phi(1,y)
=\left\{
\begin{array}{ll}
\alpha_{|y|} \hspace*{ .6 cm}
\mbox{ if $\alpha_{|y|} >0$}
\\
\\
\frac{T}{J} \log 2 \hspace*{ .4 cm} \mbox{ if $\alpha_{|y|}<0$}
\end{array}
\right.
\EEQ
The analogue of Eq. (\ref{T0l}) is, in this case,
\BEA
&&\beta J\int_{-\infty}^{\infty}dy~P(1,y)
\left[\phi(1,y)-y~m(1,y)\right]
\label{T01}
\\
\nn
&&
\hspace*{2  cm}= (1-\overline{\rho})~\log 2 + \overline{\rho}~\Theta_1(0) 
+O(e^{-1/T})
\EEA
Since the zero temperature 
density is less than one, there will be a fraction 
$1-\overline{\rho}$ of spins whose orientation is irrelevant for measuring any 
observable. This  brings to a degeneracy in the ground state, of
$2^{N(1-\overline{\rho})}$ equivalent configurations. 
Then, for the entropy density, we get
\BEA
s&=&\lim_{\beta\to\infty}
\left\{-\frac{(\beta J)^2}{4}\left[\rho-q(1)\right]^2
\right.
\\
\nn
&&\hspace*{1 cm}
\left.+(\overline{\rho}-\rho)\Theta_1+(1-\overline{\rho})\log 2\right\}
\\ \nn
&&
= (1-\overline{\rho})\log 2.
\EEA

Also in this case we can see from numerical data at low temperature that
the behavior of $\rho- q(1)$ and $s$ is $T^2$, whereas  
${\overline{\rho}}-\rho\sim T^3$ is still consistent with our numerical datas.

At zero temperature the free and internal energy are:
\BEQ
u=f=-\frac{K}{2}\rho-\mu~\rho+\int_0^1dx~q(x)~\dot\Delta(x)
\EEQ
where $\rho$ is either $1$ or $\overline{\rho}$ depending on the sign of
$\Theta_1$.

%%%%%%%%%%%%%%%%%%%%%%%%%%%%%%%%%%%%%%%%%%%%%%%%%%%%%%%

	\section{Phase Diagrams}	
	     
		\label{ss:PhDi}

Analyzing the stability of the RS solution [see Eqs. (\ref{l2},\ref{l1})]
 one gets
the critical lines
\begin{eqnarray}
&&1-(\beta J\rho)^2=0 \ ,
\label{eq:L0}
\\
&&1-\beta\tilde{K}(1-\rho)\rho=0 \ ,
\label{eq:L1}
\end{eqnarray}
above which
the only solution is the PM solution $q(x)\equiv 0$
for $x\in[0,1]$,
$\rho = 1 / [ 1 + e^{-\Theta_1}]$ [see Eq. (\ref{f:621f})], stable for 
any value of $K$. In the $T-\rho$ plane, these are, 
respectively, the straight line and the 
left branch of the spinodal line shown in Fig.
\ref{fig:phd.K1.0_Tr}, for $K=J$.
% \ref{fig:phd.K0.0_Tr} and \ref{fig:phd.K-1.0_Tr}
%(for , 0, -J$, respectively).
The two lines meet at the
tricritical point (\ref{f:622b})-(\ref{f:622c}).
Between them, under the two hard line curves branching out from the tricritical
point, there is a region of coexistence of phases (as indicated in the plot
of Fig. \ref{fig:phd.K1.0_Tr}).
The broken line curves are the first order transition lines. 
When $\mu<\mu_c$, in cooling at fixed
chemical potential, the density of the system jumps discontinuously
from the PM to the higher SG value. As an example the line at $\mu=-1.05$
is plotted.

\begin{figure}[t!]
\begin{center}
{
 \includegraphics*[width=0.49\textwidth]{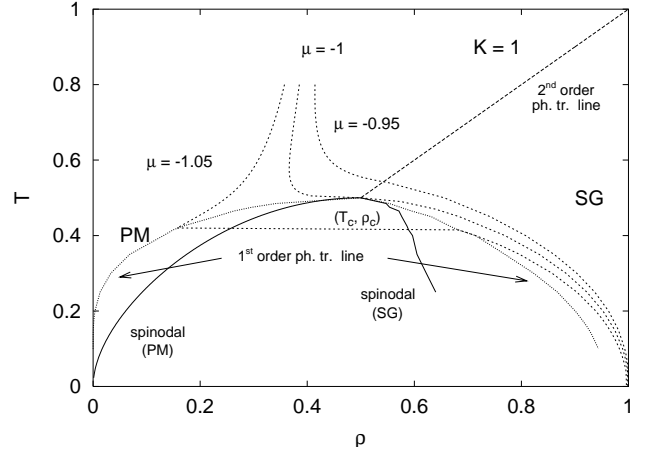}
}
\end{center}
\protect\caption{\small{
$T-\rho$ phase diagram of the RBEGC for $K/J=1$. The dot marks the
         tricritical point $\mu_c/J = -1$, $T_c=J/2$, $\rho_c=1/2$.  In
         the upper-left region we have the paramagnetic (PM) phase.
         In the upper-right region the FRSB spin glass (SG) phase.
         Following the iso-potential line at $\mu/J=-1.05$ one sees the
         jump in density at the first order phase transition.}}
\label{fig:phd.K1.0_Tr}
\end{figure}

\begin{figure}[th!]
\begin{center}
{
\includegraphics*[width=0.49\textwidth]
		 {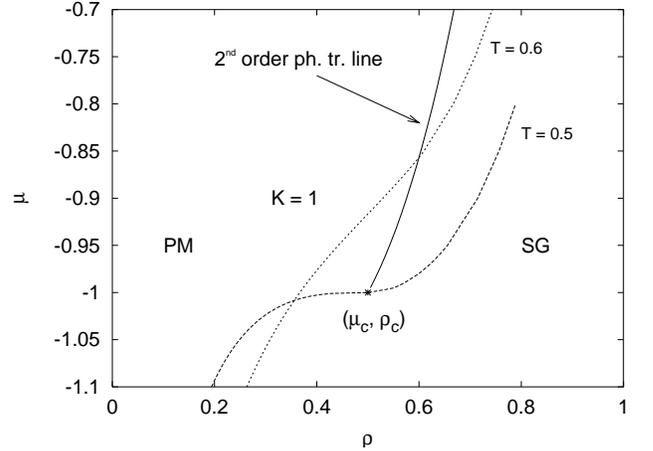}
}
\end{center}
\protect\caption{\small{$\mu-\rho$ phase diagram of the RBEGC for
$K=1$. Above  the tricritical point (star) the transition is second order.
Two isothermal
lines are shown for temperature $T=0.5, 0.6$, i.e. at and above.
}}
\label{fig:phd.K1.0_Mr1}
\end{figure}

\begin{figure}[h!]
\begin{center}
{
\includegraphics*[width=0.49\textwidth]
		 {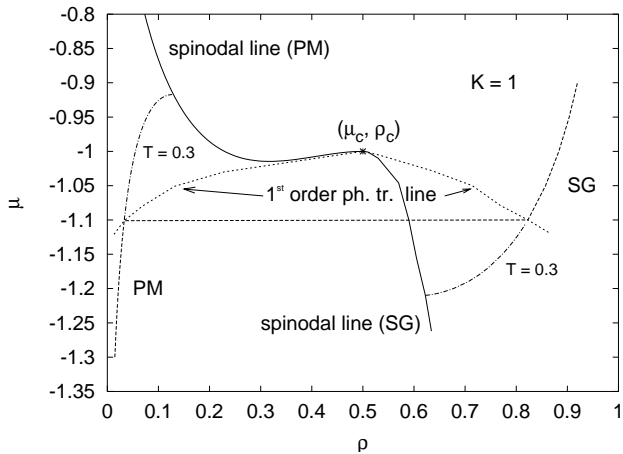}
}
\end{center}
\protect\caption{\small{$\mu-\rho$ phase diagram of the RBEGC for
$K=1$.  From the tricritical point (star) two solid lines come out: the
spinodal lines at which the paramagnetic (upper-left curve) and the
spin glass (lower-right curve) phases cease to exist, even as
metastable.  The first order transition lines, also branching out of
the tricritical point, are plotted as dashed curves.  An isothermal
line is shown, for a temperature $T=0.3$ below $T_c=1/2$.  Along such
curve a first order phase transition occurs.  In the plot, also the
metastable branches are shown, both in the RS PM phase and in the FRSB
SG phase (broken curves continuing the full curves).  They reach the
spinodal lines with zero derivative. In this plane of conjugated
thermodynamic variables a Maxwell construction can be explicitly
performed.}}
\label{fig:phd.K1.0_Mr2}
\end{figure}

\begin{figure}[th!]
\begin{center}
{
\includegraphics*[width=0.49\textwidth,height=.40\textwidth]{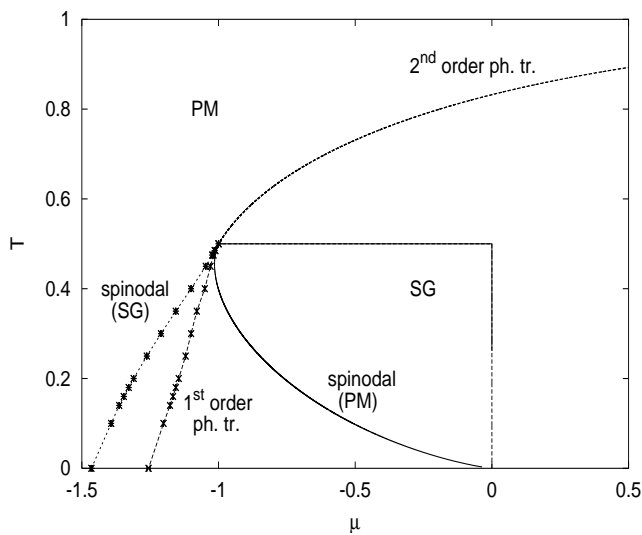}
}
\end{center}
\protect\caption{$T-\mu$ phase diagram of the RBEGC for $K=1$.
For low $\mu$ (lower than -1.2 in this specific case)
the code at $T=0$ - in the Sommers gauge - is very unstable.
Therefore the zero-$T$ transition points are  obtained by fit.}
\label{fig:phd.K1.0_TM}
\end{figure}

By crossing the critical line Eq. (\ref{eq:L0}) above the tricritical point
($\rho > \rho_c$, $T>T_c$,  $\mu> \mu_c$ )
the system undergoes a continuous phase transition
of the SK-type to a FRSB SG phase, with a non-trivial continuous 
order parameter function $q(x)$ which smoothly
grows from zero.

Below the tricritical point the scenario is completely different,
displaying
a {\it discontinuous} transition from the PM phase to
a FRSB SG phase with $q(x)$ which discontinuously 
jumps from zero to a non-trivial (continuous) function.
At the critical temperature the 
entropy is discontinuous, see Fig. \ref{fig:ent.K1.0}, 
and hence a latent heat
is involved in the transformation  from the PM to the SG phase,
implying that the transition is of the 
{\it first order} in the {\em thermodynamic} sense.
The transition line is determined by the free energy balance between
the PM and the SG phase, and is shown as a
broken line in the phase diagrams.
The line yielded by Eq. (\ref{eq:L1}) where the
PM solution becomes unstable, and the equivalent line 
from the SG side are the {\it spinodal} lines.

Also interesting is the $\mu-\rho$ phase diagram represented in
 Figs. \ref{fig:phd.K1.0_Mr1}, \ref{fig:phd.K1.0_Mr2}.  We indeed see
 that the isothermal lines cross the instability lines with 
 vanishing derivative (Fig. \ref{fig:phd.K1.0_Mr2}) and hence a
 diverging compressibility $\kappa$ occurs crossing these lines.

It can be shown that the first order transition line can be determined
in the $\mu-\rho$ phase diagram from the isothermal and spinodal lines
by using a Maxwell construction.  In the region between the first
order transition line and the spinodal line the pure phase is
metastable.  Below the spinodal lines (in the $T-\rho$ plane) no pure
phase can exist and the system is in a mixture of PM and SG phase
({\it phase coexistence}). 

Eventually, the phase diagram in the $T-\mu$
plane, for $K=1$, is shown in Fig. \ref{fig:phd.K1.0_TM}. Since our
code, even in 'most suitable' gauge (see Sec. \ref{ss:Gauge}), is
unstable in region of both low temperature and low chemical potential
we were not able to reliably compute the points belonging to the SG
spinodal line and the first order phase transition line for $T<0.1=T_c/5$.
The prolongation of the lines down to zero temperature are, hence, 
computed by fit.
Therefore the zero-$T$ transition are  estimates rougher than the others.
The first order line goes to $\mu_0^{\rm 1st}= -1.256 \pm 0.009$
 and the spinodal one
reaches zero at  $\mu_0^{\rm SG}=-1.465 \pm 0.008$.
Nevertheless, it seems that we can rule out a reentrance to the PM phase
as $T\to 0$.

By varying $K$ the scenario remains qualitatively unchanged [look at
Ref.  \cite{CLPRL02} for the phase diagrams of the Ghatak-Sherrington
model \cite{GSJPC77} ($K=0$) and the frustrated Ising lattice gas
model \cite{ANSJPF96} ($K=-J$)].  The only effect of a strong
repulsive particle-particle interaction is to increase the phase
diagram zone where the empty system ($\rho=0$) is the stable solution.
In order to find further phases, e.g. an antiquadrupolar phase,
\cite{SNAJPF97} a generalization of the present analysis to a two
component magnetic model, \cite{KSSPJETP85} including quenched
disorder, has to be carried out. \cite{CCL}

If we now look at the qualitative reproduction of the phase diagram of the 
original BEG model, with ferromagnetic interaction, in Fig. \ref{fig:BEG}
one can imagine a 
correspondence between the PM phase of the disordered magnetic material 
and the fluid phase of the He$^3$-He$^4$ mixture, and between the SG phase
and the superfluid phase. The density $\rho$ corresponds to the He$^4$ density
and the density $x$ of He$^3$ particles is $1-\rho$.
The second order transition line corresponds to the  $\lambda$ transition 
points from fluid to superfluid.

%%%%%%%%%%%%%%%%%%%%%%%%%%%%%%%%%%%%%%%%%%%%%%%%%%%%%%%

\subsection{Transition Lines}
		\label{ss:TrLi}

\begin{figure}[t!]
\begin{center}
{
\includegraphics*[width=0.49\textwidth,height=.40\textwidth]{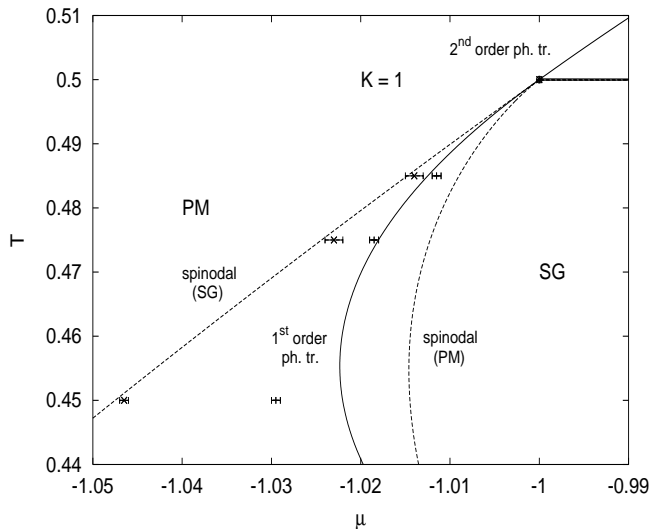}
}
\end{center}
\protect\caption{The $T$-$\mu$ phase diagram around the tricritical point.
The points displayed are those computed by directly solving the
FRSB equations at given $T$ and probing in $\mu$ where the SG
solution disappears (SG spinodal) and when the SG free energy becomes
lower than the PM one (first order transition).
The lines are obtained, instead, from the free energy 
expanded to the second  order in $\delta T$.
Also the second order transition and the PM spinodal line are displayed.
}
\label{fig:phdi_exp}
\end{figure}

The starting point to analytically determine the transition lines
around the tricritical point is the expansion of Eq. (\ref{f:62a}) for
small $q$ and densities next to $\rho_c$ that we report in appendix
\ref{app:BEGC2} [Eq. (\ref{exp})].  There the various coefficients are
expressed in terms of the function $\rho_0=(1+\exp(-\beta\mu-\beta \Kt
T/J))^{-1}$ coinciding with the paramagnetic density of the system
evaluated along the second order transition line $\Lambda_0=0$ (along
which is $\rho=\rho_0=T/J$).  The expansion parameters are the elements of
the overlap matrix $q_{ab}$ and $r\equiv\rho-\rho_0$ (replica
independent).

One can obtain the approximated analytical expression of the spinodal
line at which the FRSB solution disappears and the first order
transition line, at which the paramagnetic free energy overcomes the
approximated spin glass free energy, around the tricritical point.

In figure \ref{fig:phdi_exp} we show both the behavior of
the spinodal line of the spin
glass phase and the behavior
 of the first order transition line for the case $K=J$,
in the neighborhood of the
tricritical point  in the $T$-$\mu$ plane.
The points in the plot are  obtained by numerically solving the FRSB
differential equations.
Notice that, at the order  of approximation used,
 the first order line
displays a reentrance that does not occur, instead, 
performing the exact computation.

Notice that, at the present degree of approximation,
 already at such small distance from $(\mu_c,T_c)$ the first order line
displays a reentrance, whereas the exact computation shows that 
this does not take place. 

In the following we
consider $\mu/J \to \mu$ and $T/J\to
T$ and we express everything in terms of the small quantities
$\delta \mu\equiv \mu -\mu_c$, $\delta T\equiv
\mu -T_c$ and 
\BEQ \epsilon\equiv \delta \mu
-\frac{\p \mu_{0}(T)}{\p T} \delta T \ , 
\EEQ 
This last auxiliary variable represents the distance
 from the second order phase transition line. If 
$\epsilon < 0$, in the $T$-$\mu$ plane,  we are above the
 second order line (PM phase), otherwise below (SG phase).
The function $\mu_0(T)$ is defined as:
\BEQ
\mu_0(T)=-\frac{1}{2}-T K-T\log\left(\frac{1}{T}-1\right)
\label{mu_0}
\EEQ
The expressions of $r=\rho-\rho_0$ and
$q_1$, the highest value of $q(x)$, are 

\BEA
&&q_1=-\frac{3}{2}\delta T+\frac{1}{2}\sqrt{\epsilon-\delta T^2}
\\
%&&\nn
%\hspace*{1 cm}+ \delta T^2
%+\epsilon -\frac{\delta T}{\sqrt{\epsilon-\delta T^2}}
%\left(\delta T^2+\epsilon\right)
% \\ 
&&r=-\frac{\delta T}{2}+\frac{1}{2}\sqrt{\epsilon-\delta T^2} 
%\\
%&&\nn
%\hspace*{1cm}+ \delta T^2
%+\epsilon -\frac{\delta T}{\sqrt{\epsilon-\delta T^2}}
\EEA
Since $\delta T$ is always negative the above parameters are always positive,
as far as they exist ($\epsilon\geq \delta T^2$). The solution breaks down at
the spinodal line, therefore given by $\epsilon=\delta T^2$, i.e.
\BEQ
\mu = \mu_c - T_c + T + (T-T_c)^2
\EEQ
From the comparison of the free energy values for the PM and the SG 
phases 
the approximated first order transition line turns out to be
\BEQ
\mu =  \mu_c - T_c + T + 11.19(T-T_c)^2
\EEQ

%%%%%%%%%%%%%%%%%%%%%%%%%%%%%%%%%%%%%%%%%%%%%%%%%%%%%%%%%%%%%%
	\section{Conclusions}
     	\label{ss:sum7}

In the present paper we have shown, in some detail, the properties of the 
Random Blume-Emery-Griffiths-Capel model. 
Such a model can be seen both as a 1-spin model or a Ising-spin model
on a lattice gas.  The two formulations are completely equivalent, at
least from a static point of view. 
In the second representation, that we adopted,
the interactions involved are a  quenched random magnetic coupling $J_{ij}$
between spins and an attractive/repulsive coupling between (full) sites ($K$).
The system is embedded in a reservoir and the exchange of particles is
controlled by the chemical potential $\mu$.
The quenched random interaction is the source of a spin-glass phase at low 
temperature, provided that the chemical potential is
  large enough. If, on the contrary,
 $\mu$ is lower than a certain value, the system happen to be always
in the paramagnetic phase, becoming progressively empty as $T\to 0$.
We analytically studied the system in the mean field approximation.
As already shown by the authors in Ref.
 [\onlinecite{CLPRL02}]
the qualitative features of the system do  not depend of the value of $K$,
nor on the coupling being repulsive, attractive or zero.

  The external parameters are the temperature and the chemical
potential.  In a certain region of the phase diagram (see figure
\ref{fig:phd.K1.0_TM} in Sec. \ref{ss:PhDi}) for the $T-\mu$ diagram
at particle-particle interaction $K=J$, or, else, figures
\ref{fig:phd.K1.0_Tr}, \ref{fig:phd.K1.0_Mr1} and
\ref{fig:phd.K1.0_Mr2}) the system undergoes a second order transition
varying $T$ or $\mu$, down to a given 'tricritical' point at which the
second order line ends.  Indeed, the main feature of this model is
that for low temperature, or high chemical potential, a continuous
transition (in the Ehrenfest sense) from a pure PM phase to a pure SG
phase does not occur anymore. In its place a first order transition
takes place, with consumption/production of latent heat (see
Fig. \ref{fig:ent.K1.0} in Sec. \ref{ss:BEGCTD}) and the appearance of
a region of the parameter space where the two phases (PM and SG) do
coexist.  The SG phase comes out to be stable exclusively in the FRSB
scheme of computation. We can, thus, rule out the existence of a
glass-like phase, by this meaning a phase with one step RSB in the
static, corresponding to a dynamic decoupling of the characteristic
time scales of the processes involved in two time sectors.  Only one
kind of spin glass phase exist in the frozen phase and this is the one
typical of mean-field models for amorphous systems (i.e. spin-glasses
in the proper sense).  
In order to recover a system with a stable 1RSB
phase and displaying a first order transition a lattice gas of spins
interacting through multiple spins interaction (i.e. a $p$-spin spin
glass on lattice gas) has to be considered (see
e.g. Ref. [\onlinecite{MEPL86}]). It is interesting to notice that a
discontinuous transition between liquid and glass, 
with coexistence of phases, has been recently 
found in lattice heteropolymers with random interactions \cite{MMM1,MMM2}.
We also mention that generalizing the present model to a two component
magnetic model \cite{KSSPJETP85}, and including quenched disorder,
further phases can be found, e.g. an antiquadrupolar phase.

Computing the state of the system down to very low temperature
(including zero for not extremely low $\mu$ values)
 it has been possible to see that the reentrance displayed in the
$T$-$\mu$ phase diagram  in the  
Replica Symmetric approximation is just an artifact
(see e.g. [\onlinecite{MSJPC85}], but also 
the 'small $q$' expansion in Appendix B).
There is no $\mu$-range for which
lowering the temperature 
the  spin glass can transform itself in a paramagnet.

  We have discussed in
detail the numerical method that we use to solve the antiparabolic
differential Parisi equation
for the present model, allowing us to compute the {\em overlap}
order parameter and
all the thermodynamic observables in the SG phase. 
In particular we face the problem of making the code converge 
at very low temperature, including zero.
Section \ref{s:var} 
is dedicated to the presentation and explanation of the variational approach
to the problem of computing the FRSB anti-parabolic differential equations
  and the pseudo-spectral method
employed to solve them.

Besides the numerical resolution of the FRSB equations we have also
studied the phase diagrams around the tricritical point making an
expansion for small overlap values and densities next to the density
at the tricritical point (see Sec. \ref{ss:TrLi} and appendix
\ref{app:BEGC2}).  The expansion  formalism is valid all along, and around, the
second order transition line. The only interesting point around which
is worth probing the parameter space is, however, the tricritical
point.  From there one can build analytical approximated expressions
for the spinodal line of the spin glass phase and for the first order
transition line as well.  Above the tricritical point the transition
is always second order and qualitatively identical to the phase
transition taking place in the Sherrington-Kirkpatrick model, that is
easily recovered as limit of the present model (density one, zero
chemical potential, full lattice).

%%%%%%%%%%%%%%%%%%%%%%%%%%%%%%%%%%%%%%%%%%%%%%%%%%%%%%%
	%\appendix{}
	\addcontentsline{toc}{section}{APPENDIX A:
		Stability Analysis in the Replica Formalism}
	\section*{APPENDIX A: Stability Analysis in the Replica Formalism}
	\label{app:BEGC1}
Rescaling $\beta J \to \beta$, $J_0/J\to J$, $K/J\to K$, $h/J \to h$,
$\mu/J \to \mu$ and, consequently $\Kt\to\Kt/J$
we write the replica thermodynamic potential (\ref{f:62a}) as 
\BEA
&&G(\{\rho\},\{m\},\{q\})\equiv n\beta f(\{\rho\},\{m\},\{q\})
\label{f:C1a}
\\
\nn
&&\hspace*{1 cm}=\frac{\beta \Kt}{2}
\sum_{a=1}^{n}\rho_a^2+\frac{\beta J_0}{2}\sum_{a=1}^n 
m_a^2
+\frac{\beta^2}{4}\sum_{a\neq b} q_{ab}^2
-\log Z'
\EEA
with $Z'$ and $H'$ given respectively in Eqs. (\ref{f:62b}),
(\ref{f:62c}) and repeated here for clarity:
\BEA
&&Z'\equiv\sum_{\{S\},\{n\}}\exp\left\{-\beta H'[\{d\},\{m\},\{q\}]\right\}
\label{rep:62b}
\\
&&-\beta H'\left[\{\rho\},\{m\},\{q\}\right]\equiv\sum_a n_a
\left(\beta \Kt\rho_a +\beta \mu\right)
\label{rep:62c}
\\
\nn
&&
\hspace*{ 1 cm}+\sum_{a=1}^n S_an_a\left(\beta J_0 m_a+\beta h\right)
+\frac{\beta^2}{2}\sum_{a\neq b}q_{ab}S_an_aS_bn_b
\EEA

The variation of $G$ with respect to the replica  parameters is
\BEA
&&\delta G(\{\rho\},\{m\},\{q\}) = \beta \Kt
\sum_{a=1}^{n}\rho_a~\delta\rho_a
\label{f:C1b}
\\
\nn
&&\hspace*{.2 cm}+\beta J_0\sum_{a=1}^n  m_a~\delta m_a
+\frac{\beta^2}{2}\sum_{a\neq b} q_{ab}~\delta q_{ab}
-\sum_{a=1}^n \beta \Kt\left<n_a\right>~\delta\rho_a 
\\
\nn
&&\hspace*{.2 cm}
+\sum_{a=1}^n \beta J_0 \left<S_an_a\right> ~\delta m_a
+\frac{\beta^2}{2}\sum_{a\neq b}^{1,n}\left<S_an_aS_bn_b\right>~\delta q_{ab}
\EEA
where the average $\left<\ldots\right>$ is performed with the measure
given by Eq. (\ref{rep:62c}). From $\delta G$
the saddle point equations
 are  immediately derived:
\BEA
&&q_{ab}=\left<S_an_aS_bn_b\right>
\label{rep:62d}
\\
&&\rho_a=\left<n_a\right>
\label{rep:62e}
\\
&&m_a=\left<S_an_a\right>
\label{rep:62f}
\EEA

Eventually the fluctuations functional is equal to
\BEA
&&\delta^2 G (\{\rho\},\{m\},\{q\})=\beta \Kt
\sum_{a=1}^{n}\left(\delta\rho_a\right)^2
\label{f:C1c}
\\
\nn
&&\hspace*{.2 cm}+\beta J_0\sum_{a=1}^n  \left(\delta m_a\right)^2
+\frac{\beta^2}{2}\sum_{a\neq b} \left(\delta q_{ab}\right)^2
\\
\nn &&\hspace*{.2 cm}
-\sum_{a=1}^n \beta \Kt\delta\left(\left<n_a\right>\right)~\delta\rho_a 
+\sum_{a=1}^n \beta J_0\delta\left( \left<S_an_a\right>\right) ~\delta m_a
\\
\nn
&&
\hspace*{.2 cm}
+\frac{\beta^2}{2}\sum_{a\neq b}^{1,n}
\delta\left(\left<S_an_aS_bn_b\right>\right)~\delta q_{ab} \ ,
\EEA

\noindent  where
\BEA
&&\delta\left<o_{a_1}\ldots o_{a_k}\right>
=\delta\frac{\sum_{\{S,n\}} o_{a_1}\ldots o_{a_k} e^{-\beta H'}}{Z'}
\\
\nn
&&\hspace*{.5 cm}
=\left<o_{a_1}\ldots o_{a_k}~\delta\left(-\beta H'\right)\right>
-\left<o_{a_1}\ldots o_{a_k}\right>\left<\delta\left(-\beta H'\right)\right>
\EEA

\noindent and
\BEA
&&\delta\left(-\beta H'\right)=\beta \Kt\sum_{a=1}^n n_a~\delta\rho_a  
\\
\nn
&&\hspace*{.5 cm}
+\beta J_0 \sum_{a=1}^n S_an_a~\delta m_a
+\frac{\beta^2}{2}\sum_{a\neq b}S_an_aS_bn_b~\delta q_{ab} \ .
\EEA

In (\ref{f:C1c}) we have, thus, six kinds of terms: the quadratic ones and 
the mixed ones 
whose coefficients are given in table \ref{tab:C1a}.

The  functional $\delta^2G$, that has to be
 definite positive in order 
for the solution around which the expansion is performed
to be stable, takes the form
\BEA
\delta^2 G&=& \beta^2\sum_{ab}(\delta q_{ab})^2 
+ \beta\Kt \sum_a(\delta\rho_a)^2 
\label{f:C1d}
\\
&&+\beta J_0\sum_a(\delta m_a)^2
+\sum_{(ab),(cd)} \delta q_{ab}~ A_{(ab)(cd)}~\delta q_{cd}
\nn
\\
\nn
&&+\sum_{(ab),c}\delta q_{ab}~ D_{(ab)c}~ \delta\rho_c 
+\sum_{(ab),c}\delta q_{ab}~ E_{(ab)c}~ \delta m_c 
\\
\nn
&&+\sum_{a,(cd)}\delta\rho_a~ D_{a(cd)} ~\delta q_{cd} 
+\sum_{a,c}\delta\rho_a~B_{ac}~\delta\rho_c 
\\
\nn
&&+\sum_{a,c}\delta\rho_a~ F_{ac} ~\delta m_c
+\sum_{a,(cd)}\delta m_a~ E_{a(cd)} ~\delta q_{cd} 
\\
\nn
&&
+\sum_{a,c}\delta m_a~ F_{ac} ~\delta\rho_c 
+\sum_{a,c}\delta m_a ~ C_{ac} ~\delta m_c \ ,
\EEA

and the eigenvalues equations, then, come out to be
\BEA
&&\beta^2~ \delta q_{ab} + \sum_{(ab),(cd)}A_{(ab)(cd)}~\delta q_{cd}
\label{f:C1e}
\\
\nn
&&\hspace*{1cm}+\sum_{c}D_{(ab)c} ~\delta \rho_{c} 
+\sum_{c}E_{(ab)c} ~\delta m_{c} = \Lambda ~\delta q_{ab} \ ,
\\
&&\beta\Kt~\delta \rho_a+\sum_{c,d} D_{a(cd)} ~\delta q_{cd} 
\label{f:C1f}
\\
\nn
&&\hspace*{1cm}+\sum_c B_{ac}~\delta\rho_c 
+\sum_c F_{ac} ~\delta m_c= \Lambda~ \delta\rho_a \ ,
\\
&&\beta J_0 ~\delta m_a +\sum_{c,d} E_{a(cd)} ~\delta q_{cd} 
\label{f:C1g}
\\
\nn
&&\hspace*{1cm}+\sum_c F_{ac} ~\delta\rho_c 
+\sum_c C_{ac} ~\delta m_c= \Lambda ~\delta m_a \ .
\EEA

\begin{widetext}
\begin{center}
\begin{table}[h!]
\begin{tabular}{|c|c|}
 \hline   & {\vspace*{-3 mm}}
\\
 & {\large{ Second order expansion term $\delta^2 G$}}
\\ 
& {\vspace*{-3 mm}}
\\
 \hline   & {\vspace*{-3 mm}}
\\
{{ fluctuation term}} &
{{ coefficient}} 
\\ 
& {\vspace*{-3 mm}}
\\
\hline  & {\vspace*{-3 mm}}
\\
$\delta q_{ab}~\delta q_{cd}$ & $A_{(ab)(cd)} = $
$-\beta^4\left(\left<n_aS_an_bS_bn_cS_cn_dS_d\right>
-\left<n_aS_an_bS_b\right>\left<n_cS_cn_dS_d\right>\right)$
\\ 
& {\vspace*{-3 mm}}
\\
 \hline & {\vspace*{-3 mm}} 
\\
$\delta\rho_a~\delta\rho_c$&  $B_{ac} = $
$-(\beta\Kt)^2\left(\left<n_an_c\right>
-\left<n_a\right>\left<n_c\right>\right)$
\\ 
& {\vspace*{-3 mm}}
\\
 \hline & {\vspace*{-3 mm}} 
\\
$\delta m_a~\delta m_c$& $C_{ac} = $
$-(\beta J_0)^2\left(\left<n_aS_an_cS_c\right>
-\left<n_aS_a\right>\left<n_cS_c\right>\right)$
\\  
& {\vspace*{-3 mm}} 
\\
\hline & {\vspace*{-3 mm}} 
\\
$\delta q_{ab}~\delta \rho_{c}$ & $D_{(ab)c} = $
$-\beta^3\Kt\left(\left<n_aS_an_bS_bn_c\right>
-\left<n_aS_an_bS_b\right>\left<n_c\right>\right)$
\\ 
& {\vspace*{-3 mm}}
\\
 \hline & {\vspace*{-3 mm}} 
\\
$\delta q_{ab}~\delta m_c$&  $E_{(ab)c} = $
$-\beta^3 J_0\left(\left<n_aS_an_bS_bn_cS_c\right>
-\left<n_aS_an_bS_b\right>\left<n_cS_c\right>\right)$
\\ 
& {\vspace*{-3 mm}}
\\
 \hline & {\vspace*{-3 mm}} 
\\
$\delta \rho_a~\delta m_c$& $C_{ac} = $
$-\beta^2 \Kt J_0\left(\left<n_an_cS_c\right>
-\left<n_a\right>\left<n_cS_c\right>\right)$
\\  
& 
\\
\hline
\end{tabular}
\protect \caption{\small{Different contributions to the fluctuation functional
of the Random BEGC model in the replica formalism.  
All the averages have to be evaluated at the saddle point given
 by Eqs. (\ref{f:62d})-(\ref{f:62e}).
The subscript $(ab)$ means {\em distinct} pairs of indexes.}}
\label{tab:C1a}
\end{table}
\vspace*{-.0 cm}
\begin{table}[h!]
\begin{tabular}{|l|c|r|}
 \hline  & {\vspace*{-3 mm}}  &
\\
 &{\large{Replica Symmetric Expressions for the Coefficients in  $\delta^2G$}}&
\\
\hline   & {\vspace*{-3 mm}}  &
\\
$A_{(ab)(ab)}=$ & 
	$-\beta^4\bigl[\left<n_an_b\right>
	-\left<n_aS_an_bS_b\right>^2\bigr]$ 
		= 
	$-\beta^4\bigl[\int\D y~\tilde\rho^2(y)
	-\left(\int\D y~\tilde m^2(y)\right)^2\bigr]$
		& 
	$=A_2$
\\
& {\vspace*{-3 mm}}  &
\\
\hline   & {\vspace*{-3 mm}}  & 
\\
$A_{(ab)(ad)}=$ &
	$-\beta^4\bigl[\left<n_an_bS_bn_cS_c\right>
	-\left<n_aS_an_bS_b\right>\left<n_aS_an_cS_c\right>\bigr]$ 
 		=
	$-\beta^4\bigl[\int\D y~\tilde\rho(y)~\tilde m^2(y)
	-\left(\int\D y~\tilde m^2(y)\right)^2\bigr]$ 
		&
	$=A_1$
\\
& {\vspace*{-3 mm}} & 
\\
 \hline   & {\vspace*{-3 mm}} & 
\\
$A_{(ab)(cd)}=$ & 
	$-\beta^4\bigl[\left<n_aS_an_bS_bn_cS_cn_dS_d\right>
	-\left<n_aS_an_bS_b\right>\left<n_cS_cn_dS_d\right>\bigr]$ 
		=
	$-\beta^4\bigl[\int\D y~\tilde m^4(y)
	-\left(\int\D y~\tilde m^2(y)\right)^2\bigr]$
		&
	$=A_0$
\\
& {\vspace*{-3 mm}} & 
\\
\hline & {\vspace*{-4 mm}} & 
\\
 \hline   & {\vspace*{-3 mm}} & 
\\
$B_{aa}=$	&
	$-\beta^2\Kt^2\bigl[\left<n_a\right>
	-\left<n_a\right>\left<n_b\right>\bigr]$
	=
	$-\beta^2\Kt^2\bigl[\int\D y~\tilde\rho(y)
	-\left(\int\D y~\tilde\rho(y)\right)^2\bigr]$
	&
	$=B_1$
\\
& {\vspace*{-3 mm}} & 
\\
\hline   & {\vspace*{-3 mm}} & 
\\
$B_{ac}=$	&
	$-\beta^2\Kt^2\bigl[\left<n_an_c\right>
	-\left<n_a\right>\left<n_c\right>\bigr]$
	=
	$-\beta^2\Kt^2\bigl[\int\D y~\tilde\rho^2(y)
	-\left(\int\D y~\tilde\rho(y)\right)^2\bigr]$
	&
	$=B_0$
\\
& {\vspace*{-3 mm}} & 
\\
\hline & {\vspace*{-4 mm}} & 
\\
 \hline   & {\vspace*{-3 mm}} & 
\\
$C_{aa}=$	&
	$-\beta^2J_0^2\bigl[\left<n_a\right>
	-\left<n_aS_a\right>^2\bigr]$
	=
	$-\beta^2J_0^2\bigl[\int\D y~\tilde\rho(y)
	-\left(\int\D y~\tilde m(y)\right)^2\bigr]$
	&
	$=C_1$
 \\
 & {\vspace*{-3 mm}} & 
\\
\hline   & {\vspace*{-3 mm}} & 
\\
$C_{ac}=$	&
	$-\beta^2J_0^2\bigl[\left<n_aS_an_cS_c\right>
	-\left<n_aS_a\right>\left<n_cS_c\right>\bigr]$
	=
	$-\beta^2J_0^2\bigl[\int\D y~\tilde\rho^2(y)
	-\left(\int\D y~\tilde m(y)\right)^2\bigr]$
	&
	$=C_0$
 \\
 & {\vspace*{-3 mm}} & 
\\
\hline & {\vspace*{-4 mm}} & 
\\
\hline   & {\vspace*{-3 mm}} & 
\\
$D_{(ab)a}=$	&
	$-\beta^2\Kt\bigl[\left<n_aS_an_bS_b\right>
	-\left<n_aS_an_bS_b\right>\left<n_a\right>\bigr]$
	=
	$-\beta^2\Kt\bigl[\int\D y~\tilde m^2(y)
	-\int\D y~\tilde m^2(y)\int\D y~\tilde\rho(y)\bigr]$
	&
	$=D_1$
 \\
 & {\vspace*{-3 mm}} & 
\\
\hline   & {\vspace*{-3 mm}} & 
\\
$D_{(ab)c}=$	&
	$-\beta^2\Kt\bigl[\left<n_aS_an_bS_bn_c\right>
	-\left<n_aS_an_bS_b\right>\left<n_c\right>\bigr]$
	=
	$-\beta^2\Kt\bigl[\int\D y~\tilde\rho(y)~\tilde   m^2(y)
	-\int\D y~\tilde m^2(y)\int\D y~\tilde\rho(y)\bigr]$
	&
	$=D_0$
\\
 & {\vspace*{-3 mm}} &
 \\
\hline &  {\vspace*{-4 mm}} & 
\\
\hline   & {\vspace*{-3 mm}} & 
\\
$E_{(ab)a}=$	&
	$-\beta^3J_0\bigl[\left<n_an_bS_b\right>
	-\left<n_aS_an_bS_b\right>\left<n_aS_a\right>\bigr]$
	=
	$-\beta^3J_0\bigl[\int\D y~\tilde\rho(y)~\tilde  m(y)
	-\int\D y~\tilde m^2(y)\int\D y~\tilde m(y)\bigr]$
	&
	$=E_1$
\\
 & {\vspace*{-3 mm}} &
\\
\hline   & {\vspace*{-3 mm}} & 
\\
$E_{(ab)c}=$	&
	$-\beta^3J_0\bigl[\left<n_an_bS_bn_cS_c\right>
	-\left<n_aS_an_bS_b\right>\left<n_cS_c\right>\bigr]$
	=
	$-\beta^3J_0\bigl[\int\D y~\tilde  m^3(y)
	-\int\D y~\tilde m^2(y)\int\D y~\tilde m(y)\bigr]$
	&
	$=E_0$
\\
 & {\vspace*{-3 mm}} &
\\
\hline & {\vspace*{-4 mm}} & 
\\
\hline   & {\vspace*{-3 mm}} & 
\\
$F_{aa}=$	&
	$-\beta^2J_0\Kt\bigl[\left<n_aS_a\right>
	-\left<n_a\right>\left<n_aS_a\right>\bigr]$
	=
	$-\beta^2J_0\Kt\bigl[\int\D y~\tilde  m(y)
	-\int\D y~\tilde\rho(y)\int\D y~\tilde m(y)\bigr]$
	&
	$=F_1$
\\
 & {\vspace*{-3 mm}} &
\\
\hline   & {\vspace*{-3 mm}} & 
\\
$F_{ac}=$	&
	$-\beta^2J_0\Kt\bigl[\left<n_an_cS_c\right>
	-\left<n_a\right>\left<n_cS_c\right>\bigr]$
	=
	$-\beta^2J_0\Kt\bigl[\int\D y~\tilde\rho(y)~\tilde  m(y)
	-\int\D y~\tilde\rho(y)\int\D y~\tilde m(y)\bigr]$
	&
	$=F_0$
\\
 & &
\\
\hline
\end{tabular}
\protect\caption{\small{Here we report
the coefficients of the second order term in the expansion
 of the free energy functional (\ref{f:C1a}) of the Random BEGC model
around the RS solution. The left hand side of equalities shows the 
generic expression of the coefficients, whereas the right hand side 
gives the coefficients evaluated with the {\em Ansatz} of replica symmetry.
On the far right column the abbreviation for the RS coefficients 
are listed.}}
\label{tab:C1b}
\end{table}
\end{center}
\end{widetext}

\vskip - 2cm

%%%%%%%%%%%%%%%%%%%%%%%%%%%%%%%%%%%%%%%%%%%%%%%%%%%%%%%
\addcontentsline{toc}{subsection}{A.1 \ \ The Stability of the RS Solution}
\subsection*{A.1 \ \ The Stability of the RS Solution}
Inserting in the above  expressions
the {\em Ansatz} that all replicas are equivalent (same density, same magnetization 
and same overlap between all of them) we can study the 
stability of the RS solution.
The RS {\em Ansatz} is implemented by the substitutions:
\BEQ
q_{ab}=(1-\delta_{ab})~q_0 \ \ ; \ \ 
\rho_a=\rho \ \ ; \ \ 
m_a = m \ .
\EEQ

Using the definition 
\BEQ
\Theta_0=  \beta \Kt \rho + \beta \mu -\frac{\beta~q_0}{2} \ ,
\label{rep:621b}
\EEQ
the replica one site Hamiltonian (\ref{rep:62c}) becomes:
\BEA
&&-\beta H'\left(\rho,m,q_0\right)=\Theta_0 \sum_a n_a
\label{f:C1h}
\\
\nn
&&\hspace*{ 1cm}
+\beta \left(J_0 m + h\right)\sum_a n_as_a
+\frac{\beta^2 q_0}{2}\left(\sum_an_aS_a\right)^2 \ .
\EEA

With this measure we have to compute the averages 
$\left<o_{a_1}\ldots o_{a_k}\right>$ occurring
in the saddle point equations (\ref{f:62c})-(\ref{f:62e})
 and in the coefficients of 
the eigenvalues equations (see Tab. \ref{tab:C1a}).

``Opening'', with the Hubbard-Stratonovic method,
the squared sum in Eq. (\ref{f:C1h}) 
into the Gaussian integral of an exponential with linear exponent
we get
\BEQ
e^{-\beta H'}=
\int_{-\infty}^{\infty} \D y~ \exp\left[-\sum_a H_a(y)\right] \ ,
\EEQ
with the {\em  one-index} Hamiltonian
\BEQ
H_a(y)\equiv \Theta_0 n_a+ \h(y)~ n_aS_a
\label{f:C1i}
\EEQ
and 
\BEQ
\h(y)\equiv y~\frac{\beta\sqrt{q_0}}{2}+ J_0 m+ h \ .
\label{f:C1l}
\EEQ
We also define the one-index partition sum as
\BEQ
Z_a(y)\equiv\sum_{S_a,n_a}e^{-\beta H_a(y)}=2+2~e^{\Theta_0}~\cosh \h(y) \ ,
\label{f:C1m}
\EEQ
and the one-index average
\BEQ
\left<o_a\right>_a=\frac{\sum_{S_a,n_a} o_a e^{-\beta H_a(y)}}{Z_c(y)} \ ,
\EEQ
that is  a function of $y$.

With the help of Eqs. (\ref{f:C1h}), (\ref{f:C1i}) we are now able to compute 
the nominator of the average $\left<o_{a_1}\ldots o_{a_k}\right>$:
\BEA
&&\hspace*{-.3cm}\sum_{\{S\},\{n\}} o_{a_1}\ldots o_{a_k} e^{-\beta H'}
\\
\nn
&&=\int_{-\infty}^{\infty}\hspace*{-.3cm} \D y~ 
\prod_{i=1}^k \sum_{S_{a_i},n_{a_i}}~o_{a_1}\hspace*{-.3cm}
 e^{-\beta H_{a_i}(y)}
\prod_{i=k+1}^n \sum_{S_{a_i},n_{a_i}}\hspace*{-.3cm}e^{-\beta H_{a_i}(y)}
\nn
\\
&&=\int_{-\infty}^{\infty}\hspace*{-.3cm} \D y~
\left[Z_1(y)\right]^n \prod_{i=1}^k \left<o_{a_i}\right>_{a_i} \ ,
\EEA
so that the complete expression of the average is eventually given by:
\BEA
\lim_{n\to 0}\left<o_{a_1}\ldots o_{a_k}\right>&=& 
\lim_{n\to 0}\frac{1}{Z'}\int_{-\infty}^{\infty}\hspace*{-.3cm} \D y~
\left[Z_1(y)\right]^n \prod_{i=1}^k \left<o_{a_i}\right>_{a_i} 
\nn
\\
&&=
\int_{-\infty}^{\infty} \D y~\prod_{i=1}^k \left<o_{a_i}\right>_{a_i} \ ,
\EEA
since
$Z'=\int_{-\infty}^{\infty} \D y~\left[Z_1(y)\right]^n \to 1$ in the 
zero replicas  limit.

In our case $o_a$ can be any combination of $n_a$ and $n_aS_a$ 
occurring in Eqs. (\ref{rep:62d})-(\ref{rep:62f}) and in the coefficients
of the second order expansion term of the thermodynamic potential 
[Eq. (\ref{f:C1c})] reported in Tab. \ref{tab:C1a}.
In Eq. (\ref{f:C1i}) $n_a$ is coupled to $\Theta_0$, whereas $n_aS_a$
is coupled to $\h$. Their one-index averages can, then, be
obtained as:
\BEA
\left<n_a\right>_a&=&\frac{\p\log Z_a(y)}{\p\Theta_0}
=\frac{\cosh \h(y)}{e^{-\Theta_0}+\cosh\h(y)}
\equiv\tilde\rho(y) \ ,
\nn
\\
\\
\left<n_aS_a\right>_a&=&\frac{\p\log Z_a(y)}{\p\h}
=\frac{\sinh \h(y)}{e^{-\Theta_0}+\cosh\h(y)}
\equiv\tilde m (y) \ .
\nn
\\
\EEA
Using these results we get Eqs. (\ref{f:621a})-(\ref{f:621c})
and the expression of the coefficients of $\delta^2G$ for the RS
solution.
We report the complete list  in table \ref{tab:C1b}.
Filling in those terms, the sum 
occurring in the eigenvalues Eqs. (\ref{f:C1e})-(\ref{f:C1g})
can, thus, be written in the RS scheme as:
\BEA
&&\hspace*{0.0 cm}\sum_{(cd)}A_{(ab)(cd)}~\delta q_{cd}
=(A_2-2A_1+A_0)~\delta q_{ab}
\label{f:C1p1}
\\
\nn
&&\hspace*{2 cm}	+(A_1-A_0)\sum_c(\delta q_{ac}+\delta q_{bc})
	+\frac{A_0}{2}\sum_{cd}\delta q_{cd}
\\
&&\hspace*{0.0 cm}\sum_{c}D_{(ab)c}~\delta \rho_{c}=(D_1-D_0)(\delta\rho_{a}+\delta\rho_b)
	+D_0\sum_{c}\delta \rho_{c}
\nn
\\
\label{f:C1p2}
\\
\nn
&&\hspace*{0.0 cm}\sum_{c}E_{(ab)c}~\delta m_{c}
=(E_1-E_0)(\delta m_{a}+\delta m_b)
	+E_0\sum_{c}\delta m_{c}
\\
\label{f:C1p3}
\\
&&\hspace*{0.0 cm}\sum_{(cd)}D_{a(cd)}~\delta q_{cd}
=(D_1-D_0)\sum_c\delta q_{ac}
	+\frac{D_0}{2}\sum_{cd}\delta q_{cd}
\nn
\\
\label{f:C1p4}
\\
&&\hspace*{0.0 cm}\sum_{c}B_{ac}~\delta \rho_{c}=(B_1-B_0)\delta\rho_{a}
	+B_0\sum_{c}\delta \rho_{c}
\label{f:C1p5}
\\
&&\hspace*{0.0 cm}\sum_{c}F_{ac}~\delta m_{c}=(F_1-F_0)\delta m_{a}
	+F_0\sum_{c}\delta m_{c}
\label{f:C1p6}
\\
&&\hspace*{0.0 cm}\sum_{(cd)}E_{a(cd)}~\delta q_{cd}
=(E_1-E_0)\sum_c\delta q_{ac}
	+\frac{E_0}{2}\sum_{cd}\delta q_{cd}
\nn
\\
\label{f:C1p7}
\\
&&\hspace*{0.0 cm}\sum_{c}F_{ac}~\delta \rho_{c}=(F_1-F_0)\delta\rho_{a}
	+F_0\sum_{c}\delta \rho_{c}
\label{f:C1p8}
\\
&&\hspace*{0.0 cm}\sum_{c}C_{ac}~\delta m_{c}=(C_1-C_0)\delta m_{a}
	+C_0\sum_{c}\delta m_{c}
\label{f:C1p9}
\EEA

In the case $J_0=0$ (no ferromagnetic phase) 
$C_{ab}=E_{(ab)c}=F_{ab}=0$, $\forall a,b,c$ and we are left with
a system of equations for 
$n(n-1)/2$ variables $\delta q_{ab}$ and $n$ variables $\delta\rho_a$
($\delta m_a$ are not involved, since always coupled with $J_0$).
In total,  the dimension of the space of solutions is 
$d_{\rm tot}\equiv n(n+1)/2$.

\begin{itemize}
\item{$\Lambda_0$}
\\
To obtain the first eigenvalue $\Lambda_0$ 
we analyze the equations (\ref{f:C1e})-(\ref{f:C1f}) in the subspace
\BEQ
\sum_a\delta q_{ab}=0 \ ,~\forall b \hspace*{.2 cm} ; \hspace*{1 cm} 
\delta \rho_a = 0 \ ,~\forall b \ .
\label{f:C1q}
\EEQ
These are  $2n$ equations and the dimension of this subspace is 
$d_{\Lambda_0}=d_{\rm tot}- 2 n=n(n-3)/2$,
 corresponding to the degeneracy of $\Lambda_0$.
Eq. (\ref{f:C1q}) also implies that 
$\sum_{ab}\delta q_{ab}=\sum_a \delta\rho_a=0$.

Using the conditions  (\ref{f:C1q}), the equations
 (\ref{f:C1e}), (\ref{f:C1f}) are easily reduced, in this case, to
the single equation
\BEQ
(\beta^2+A_2-2A_1+A_0)~\delta q_{ab}=\Lambda_0~ \delta q_{ab} \ ,
\EEQ
yielding
\BEQ
\Lambda_0=\beta^2+A_2-2A_1+A_0 \ .
\label{f:C1r}
\EEQ

\item{$\Lambda_1$}
\\
We look at the second eigenvalue in the subspace 
\BEQ
 \sum_{ab}\delta q_{ab}=0 \hspace*{0.5 cm} ; \hspace*{0.5 cm}
\sum_a \delta\rho_a=0 \ ,
\label{f:C1s}
\EEQ
where, as opposed to the previous case,
 $\sum_a\delta q_{ab}\neq 0$, $\forall b$ and 
$\delta_a\rho_a\neq 0$,  $\forall a$. The $\Lambda_0$ and the $\Lambda_1$
subspaces are orthogonal.

Subtracting two [the  equalities given in Eq.
(\ref{f:C1s})]  and the degeneracy  of the $\Lambda_0$
to the total dimension
 we get the degeneracy of $\Lambda_1$:
$d_{\Lambda_1}=d_{\rm tot}-d_{\Lambda_0}-2=2(n-1)$.

Eqs. (\ref{f:C1e})-(\ref{f:C1f}) reduce in this subspace to
\BEA
&&\left[\beta^2+A_2+(n-4)A_1-(n-3)A_0\right]\sum_b\delta q_{ab}
\\
\nn
&&\hspace*{1 cm}+(n-2)(D_1-D_0)~\delta\rho_a
=\Lambda_1\sum_b\delta q_{ab}
\\
&&(D_1-D_0)\sum_b\delta q_{ab}
\\
\nn 
&&\hspace*{1 cm}+
\left(\beta\Kt+B_1-B_0\right)\delta\rho_a=
\Lambda_1~~\delta\rho_a
\EEA

for which the eigenvalue comes out to be
\begin{widetext}
\BEA
\Lambda_1&=&\frac{1}{2}\left[
\beta^2+A_2+(n-4)A_1-(n-3)A_0+\beta\Kt+B_1-B_0
\right]
\label{f:C1t}
\\
\nn
&&\hspace*{2 cm}\pm\frac{1}{2}\sqrt{
\left[\beta^2+A_2+(n-4)A_1-(n-3)A_0-(\beta\Kt+B_1-B_0)\right]^2
+4(n-2)(D_1-D_0)^2}
\EEA
\end{widetext}

\item{$\Lambda_2$}
\\
To find $\Lambda_2$ we are left with  the subspace of solutions 
such that
\BEQ
\sum_{ab}\delta q_{ab}\neq 0\hspace*{0.5 cm} ; \hspace*{0.5 cm}
\sum_a\delta\rho_a\neq0 \ .
\EEQ
and whose dimension is $2$.

The eigenvalues equations become now
\BEA
&&
\left[\beta^2+A_2+(n-2)\left(2A_1+\frac{n-3}{2}A_0\right)\right]
\sum_{ab}\delta q_{ab}
\nn
\\
\nn
&&+
(n-1)\left[2D_1+(n-2)D_0\right]\sum_a\delta\rho_a=
\Lambda_2\sum_{ab}\delta q_{ab}
\\
\\
&&
\left(D_1+\frac{n-2}{2}D_0\right)\sum_{ab}\delta q_{ab}
\\
\nn
&&+
\left(\beta\Kt+B_1+(n-1)B_0\right)\sum_a\delta\rho_a
=
 \Lambda_2\sum_a\delta\rho_a
\EEA

so that it, finally,  holds
\begin{widetext}
\BEA
&&
\Lambda_2=\frac{1}{2}\left[\beta^2+A_2+2(n-2)A_1+\frac{(n-2)(n-3)}{2}A_0
+D_1+\frac{n-2}{2}D_0\right]
\label{f:C1u}
\\
\nn
&&\hspace*{1.5cm}
\pm\frac{1}{2}\sqrt{\left[\beta^2+A_2+(n-2)\left(2 A_1+\frac{n-3}{2}A_0\right)
-\left(D_1+\frac{n-2}{2}D_0\right)\right]^2+
2(n-1)\left[2D_1+(n-2)D_0\right]^2}
\EEA
\end{widetext}

\end{itemize}

\begin{widetext}
In the limit for $n\to 0$ the eigenvalues 
$\Lambda_1$ (\ref{f:C1t}) and $\Lambda_2$ (\ref{f:C1u}) are
  degenerate. They  reduce both to 
\BEA
&&\Lambda_1=\Lambda_2=\frac{1}{2}\left(
\beta^2+A_2-4A_1+3A_0+\beta\Kt+B_1-B_0\right)
\label{f:C1x}
\\
&&\hspace*{ 4cm}\pm\frac{1}{2}~\sqrt{
\left(\beta^2+A_2-4A_1+3A_0-\beta\Kt-B_1+B_0\right)^2-8(D_1-D_0)^2} \ .
\nn
\EEA
\end{widetext}

If we also put $h=0$ we simplify things much, since in 
this case the stable solution is $q_0=0$.
This brings to
\BEA
&&\tilde\rho(y)=\frac{1}{e^{-\Theta_0}+1}\equiv\rho_0 \ ,
\\
&&\tilde m(y)=0 \ ,
\EEA
and, therefore,  no integral in the coefficients of table \ref{tab:C1b}
has to be carried out anymore.  This leads to
$A_1=A_0=D_1=D_0=B_0=0$ and
\BEA
&&A_2=-\beta^4\rho_0^2 \ ,
\\
&&B_1=-\beta^2\Kt^2\rho_0(1-\rho_0) \ .
\EEA

Substituting these coefficients into Eqs. (\ref{f:C1r})-(\ref{f:C1x})
we obtain
\BEA
&&\Lambda_0=\beta^2(1-\beta^2\rho_0^2) \ ,
\\
&&\Lambda_1=\beta\Kt\left[1-\beta\Kt\rho_0(1-\rho_0)\right] \ .
\EEA
The above analysis is valid for $\Kt>0$. This is always the case ik 
the biquadratic coupling is $K\geq 0$.
When a negative $K$ occurs, however, at high enough temperature
($T>1/(2|K|)$) $\Kt$ becomes negative.
Taking this into account and repeating the whole scheme of computation 
the final result for $\Kt<0$ is, in the case of out interest,
\BEQ
\Lambda_1=-\beta\Kt\left[1-\beta\Kt\rho_0(1-\rho_0)\right] \ .
\EEQ

%%%%%%%%%%%%%%%%%%%%%%%%%%%%%%%%%%%%%%%%%%%%%%%%%%%%%%%

	\addcontentsline{toc}{section}{APPENDIX B:
		Small q Expansion}
	\section*{APPENDIX B: Small q Expansion}
	\label{app:BEGC2}

The expansion to the fourth order in $r=\rho-\rho_0$ and $q_{ab}$ is:

\BEA
&&\beta f(\{q_{ab}\},r)=\beta f_0-\beta\Kt\left(\rho_0-\frac{T}{J}\right)r
\label{exp}
\\
\nn
&&
+\frac{\Lambda_1}{2} r^2-\frac{(\beta \Kt)^3}{6}\rho_0(1-\rho_0)(1-2\rho_0)r^3
\\
\nonumber
&&
-\frac{(\beta\Kt)^4}{24}\rho_0(1-7\rho_0+12\rho^2-6\rho^3)r^4
\\
\nn
\\
&&+\frac{\Lambda_0}{4} \frac{1}{n}\Tr ~\mathbf{q}^2
-\frac{(\beta J)^4}{2}\beta \Kt \rho_0^2(1-\rho_0)
\nn
\\
\nn
&&\hspace*{1.5 cm}\times\left[r+\frac{\beta \Kt}{2}(2-3 \rho_0)r^2\right] 
\frac{1}{n}\Tr ~\mathbf{q}^2
\\
\nn
&&-\frac{(\beta J)^6}{6}\rho^3\left[1+3\beta \Kt(1-\rho_0)r\right]
\frac{1}{n}\Tr ~\mathbf{q}^3
\\
\nn
&&-\frac{(\beta J)^8}{48}
\rho_0^2\frac{1}{n}
\left[
\sum_{\langle ab\rangle}q_{ab}^4
+6\rho_0\sum_{\langle abc\rangle}q^2_{ab}q^2_{ac}
\right.
\\
\nn
&&\hspace*{ 1 cm}
\left.+\frac{3}{2}\rho_0^2\sum_{\langle abcd\rangle}q^2_{ab}q^2_{cd}
+6\rho_0^2 \sum_{\langle abcd\rangle}q_{ab}~q_{bc}~q_{cd}~q_{da}
\right]
\EEA

where $\sum_{\langle \ldots\rangle}$ is a sum over {\em distinct}
indexes. The paramagnetic contribution is

\BEQ \beta f_0=\frac{\beta \Kt}{2}\rho_0^2+\log(1-\rho_0) \EEQ 
and
\BEQ \rho_0=\frac{1}{1+\exp\left[-\beta \mu-\beta\Kt T/J\right]}
 \EEQ
is the value of the paramagnetic density evaluated along the second
order phase transition line.

The above expansion is valid in the neighborhood of the second order
transition line, $\Lambda_0=0$, down to, and including the tricritical point
($\Lambda_0=0 \cup \Lambda_1=0$).

%%%%%%%%%%%%%%%%%%%%%%%%%%%%%%%%%%%%%%%%%%%%%%%%%%%%%%%

%\subsection{Replica Symmetric case}

%At the Replica Symmetric level we look at the behavior of $r$ and $q1=q$
%around the tricritical point. For that we need to parameters, such as, e.g.
%the distance from the tricritical value of $T$ and $\mu$:
%$\delta\mu=\mu-\mu_c$, $\delta T=T-T_c$.
%A very confortable combination of the two, regarding practical computation
%is the distance from $\Lambda_0=0$, that in the $T$-$\mu$ plane reads
%\BEQ
%\mu_0(T)=-\frac{1}{2}-T K-T\log\left(\frac{1}{T}-1\right)
%\label{mu_0}
%\EEQ
%Therefore we define
%\BEQ
%\nn
%\epsilon\equiv \delta \mu-\frac{\p\mu_0(T)}{\p T}\delta T
%\EEQ

%For a generic $K$ the approximate behavior of $r$ and $q_1$
%is, then,

%\BEA
%&&r=
%\\
%&&q_1=
%\EEA

The various terms appearing in the expansion of the free energy
functional, Eq. (\ref{exp}) are expressed in the following.
In order to compute it 
in a generic $N_B$-RSB scheme we just need the following expression for the 
form of the overlap matrix $q_{ab}$

\BEA
q_{ab}&=&-q_{N_B} \delta_{ab} +\sum_{i=1}^{i=N_B} \left( q_i-q_{i-1} \right)
\eab^{(i)} +q_0
\\
&=&\sum_{i=0}^{N_B+1} \left( q_i-q_{i-1} \right) \eab^{(i)}
\\
&=&\sum_{i=0}^{N_B} q_i\left(\eab^{(i)}-\eab^{(i+1)} \right)
\label{def:C2a}
\EEA
with $q_{N_B+1}=q_{-1}=0$, $m_{N_B+1}=1$, $m_0=n$, $\eab^{N+1}=\delta_{ab}$,
$\eab^{(0)}$ a $n\times n$ matrix with all elements equal to 1.

The matrix $\eab^{(i)}$ has $m_i/m_{i+1}$ blocks along the diagonal and its
summation rules are:
\BEA
&&\sum_{b}\eab^{(i)}\ebc^{(j)}= m_{\rm max(i,j)} \eac^{\rm min(i,j)}
\\
&&\sum_{b}\delta_{ab}\ebc^{(j)}=\eac^{(j)}
\\
&&\sum_{b}\eab^{(i)}=m_i
\\
&&\sum_{bc}\eab^{(i)}\ebc^{(j)}\ecd^{(k)}
\\
\nn
&&\hspace*{1 cm}= m_{\rm max(i,j,k)}~
m_{\rm 2^{nd}max(i,j,k)}\ead^{\rm min(i,j,k)}
\\
&&\sum_{bcd}\eab^{(i)}\ebc^{(j)}\ecd^{(k)}\ede^{(l)}
\label{def:C2b}
\\
\nn
&&= m_{\rm max(i,j,k,l)}~
m_{\rm 2^{nd}max(i,j,k,l)}~m_{\rm 3^{rd}max(i,j,k,l)}\eae^{\rm min(i,j,k,l)}
\EEA
where $\sum_{\ldots}$ is a sum over {\em all} indexes.

\subsection*{Term $\Tr {\bf q}^2=\sum_{ab}q_{ab}~q_{ba}$}

\BEA
&&\sum_{b}q_{ab}~q_{bc}=\sum_{i=0}^{N_B} q_i^2 \left( m_i \eac^{(i)} -
m_{i+1} \eac^{(i+1)} \right) \ ,
\\
&&\sum_{b}q_{ab}~q_{ba}=\sum_{i=0}^{N_B} q_i^2 \left( m_i - m_{i+1} \right) \ .
\label{f:C2c}
\EEA

In the full replica symmetry breaking limit and in the zero replicas
limit the breaking parameters become continuous between $0$ and $1$:
\BEQ
\nn
m_i\to i \ \ \  m_k-m_{i+1}\to -dx \ .
\EEQ
Moreover the same structure of Eq. (\ref{f:C2c})
holds  for any $\sum_{ab}q_{ab}^p$, thus
\BEQ
\lim_{n\to 0}\frac{1}{n}\sum_{ab}q_{ab}^p=\lim_{n\to 0}\sum_{i=0}^{N_B}
q_i^p(m_i-m_{i+1})=-\int_0^1 dx~q^p(x)
\EEQ

\begin{widetext}
\subsection*{Term $\Tr {\bf q}^3=\sum_{abc}q_{ab}~q_{bc}~q_{ca}$}
Using (\ref{def:C2a})  this trace can be written as:

\BEA
&&\sum_{bc} \left(
 \eab^{(i)}\ebc^{(j)}\eca^{(k)}    -\eab^{(i+1)}\ebc^{(j)}\eca^{(k)}
-\eab^{(i)}\ebc^{(j+1)}\eca^{(k)}  +\eab^{(i+1)}\ebc^{(j+1)}\eca^{(k)}
\right. 
\\
&&\hspace*{1 cm}\left.
-\eab^{(i)}\ebc^{(j)}\eca^{(k+1)}  +\eab^{(i+1)}\ebc^{(j)}\eca^{(k+1)}
+\eab^{(i)}\ebc^{(j+1)}\eca^{(k+1)}-\eab^{(i+1)}\ebc^{(j+1)}\eca^{(k+1)}
\right) \nn
\EEA

In table \ref{tab:C2a} we report the four possible non 
degenerate contributions coming out from this sum and their multiplicity.

In the full replica symmetry breaking limit (and for number of replicas
$n\to 0$):
\BEA
\nn
m_k\to x && m_k-m_{k+1}\to -dx \ ,
\\
\nn 
m_i\to y && m_i-m_{i+1}\to -dy\ .
\EEA
Thus the above expressions for the terms contributing to the sum reduce to
\BEA
&&B\to \int_0^1 dx~q(x)\int_0^x dy~q(y)^2 \ ,
\\
&&D\to \int_0^1 dx~x~q(x)^3 \ ,
\EEA
\noindent where we have neglected  terms of order $(dx)^2$.
Summing up, the trace comes out to be:

\BEA
&&\lim_{n\to 0}\frac{1}{n}{\rm Tr}~ {\bf q}^3= 3 B + D 
\\
\nn
&&\hspace*{1 cm}=
3\int_0^1 dx~q(x)\int_0^x dy~q(y)^2 + \int_0^1 dx~x~q(x)^3
\EEA

\begin{center}
\begin{table}[t!]
\begin{center}
\begin{tabular}{|cl|c|c|}
\hline 
& {type of sum} & {contribution} &{ multiplicity} 
\\
\hline
&A:~$i<j<k$     & $0$            & $6$
\\
\hline
&B:~$i=j<k$     & $q_k~q_i^2(m_k-m_{k+1})(m_i-m_{i+1})$ & $3$
\\
\hline 
&C:~$i<j=k$     & $0$           &$3$
\\
\hline 
&D:~$i=j=k$     & $q_k^3\left[(m_k-m_{k+1})^2-m_{k+1}(m_k-m_{k+1})\right]$ &$
1$
\\
\hline  
\end{tabular}
\end{center}
\protect\caption{Contributions to the trace of ${\bf q}^3$ for a generic number
of breakings of the replica symmetry: only two kind of 
terms are different from zero.} 
\label{tab:C2a}
\end{table}
\end{center}

%%%%%%%%%%%%%%%%%%%%%%%%%%%%%%%%%%%%%%%%%%%%%%%%%%%%%%%

\subsection*{Term $\Tr{\bf q}^4=\sum_{abcd}q_{ab}~q_{bc}~q_{cd}~q_{da}$}
Exploiting the formulation of Eq. (\ref{f:C2a}) once again we can obtain

\BEA
\label{f:C2p}
&&\sum_{bcd}q_{ab}~q_{bc}~q_{cd}~q_{da}=\sum_{i,j,k}^{0,N_B} q_i~q_j~q_k~q_l
\sum_{bcd} \left( \eab^{(i)}\ebc^{(j)}\ecd^{(k)}\eda^{(l)}    
-\eab^{(i+1)}\ebc^{(j)}\ecd^{(k)}\eda^{(l)}
-\eab^{(i)}\ebc^{(j+1)}\ecd^{(k)}\eda^{(l)}  
-\eab^{(i)}\ebc^{(j)}\ecd^{(k+1)}\eda^{(l)}
\right. \nn
\\
&&\hspace*{1 cm}\left.
-\eab^{(i)}\ebc^{(j)}\ecd^{(k)}\eda^{(l+1)}
+\eab^{(i+1)}\ebc^{(j+1)}\ecd^{(k)}\eda^{(l)}  
+\eab^{(i+1)}\ebc^{(j)}\ecd^{(k+1)}\eda^{(l)}
+\eab^{(i+1)}\ebc^{(j)}\ecd^{(k)}\eda^{(l+1)}
\right.
\nn
\\
&&\hspace*{1 cm}\left.
+\eab^{(i)}\ebc^{(j+1)}\ecd^{(k+1)}\eda^{(l)}
+ \eab^{(i)}\ebc^{(j+1)}\ecd^{(k)}\eda^{(l+1)}    
+ \eab^{(i)}\ebc^{(j)}\ecd^{(k+1)}\eda^{(l+1)}
-\eab^{(i+1)}\ebc^{(j+1)}\ecd^{(k+1)}\eda^{(l)}    
\right.
\nn
\\ 
&&\hspace*{1 cm}\left.
- \eab^{(i+1)}\ebc^{(j+1)}\ecd^{(k)}\eda^{(l+1)}    
- \eab^{(i+1)}\ebc^{(j)}\ecd^{(k+1)}\eda^{(l+1)}    
- \eab^{(i)}\ebc^{(j+1)}\ecd^{(k+1)}\eda^{(l+1)}
+ \eab^{(i+1)}\ebc^{(j+1)}\ecd^{(k+1)}\eda^{(l+1)}    
\right)\nn
\EEA

\begin{center}
\begin{table}[t!]
\begin{center}
\begin{tabular}{|cl|c|c|}
\hline 
& {type of sum} & {contribution} &{multiplicity} 
\\
\hline
&A:~$i<j<k<l$   & $0$            &$24$
\\
\hline
&B:~$i=j<k<l$   & $q_l~q_k~q_i^2(m_l-m_{l+1})(m_k-m_{k+1})(m_i-m_{i+1})$     
   & $12$
\\
\hline 
&C:~$i<j=k<l$   & $0$           &$12$
\\
\hline 
&D:~$i<j<k=l$   & $0$           &$12$
\\
\hline 
&E:~$i=j<k=l$   & $q_i^2q_k^2(m_k-m_{k+1})^2(m_i-m_{i+1})$ &$6$
\\
\hline 
&F:~$i=j=k<l$   & $q_l~q_k^3(m_l-m_{l+1})\left[(m_k-m_{k+1})^2-m_{k+1}(m_k-m_
{k+1})\right]$ &$4$
\\
\hline 
&G:~$i<j=k=l$   & $0$ &$4$
\\
\hline 
&H:~$i=j=k=l$   & $q_i^4(m_i-m_{i+1})\left[m_{i+1}^2-m_{i+1}(m_i-m_{i+1})+(m_
i-m_{i+1}
)^2\right]$ &$1$
\\
\hline  
\end{tabular}
\end{center}
\protect\caption{Contributions and multiplicities of the terms in the sum 
of Eq. (\ref{f:C2p}). Only for kinds of terms are non-zero.
Of them, term $E$ turns to be of order  $(dx)^2$ once the zero replica
limit has been  performed. }
\label{tab:C2b}
\end{table}
\end{center}
\end{widetext}

Contribution and multiplicity are reported in table \ref{tab:C2b}.

For $N_B\to\infty$ and $n\to 0$ the terms are
\BEA
&&B\to -\int_0^1 dx~q(x)\int_0^x dy~q(y) \int_0^y dz q^2(z)
\\
&&E \to 0
\\
&&F \to -\int_0^1 dx~q(x)\int_0^x dy~y~q^3(y)
\\
&&H \to -\int_0^1 dx~x^2 q^4(x)
\EEA 

\subsection*{Term $\sum_{a b c}q_{ab}^2q_{ac}^2$}
Eventually this last object reads:

\BEA
&&\sum_{b c}q_{ab}^2q_{ac}^2=\sum_{b} q_{ab}^2\sum_c q_{ac}^2
\\ 
\nn
&&
=\sum_iq_i^2(m_i-m_{i+1})\sum_jq_j^2(m_j-m_{j+1}) \ ;
\EEA
the FRSB limit (for $n\to 0$) being
\BEA
&&\lim_{n\to 0}\frac{1}{n}{\sum_{abc}} q_{ab}^2q_{ac}^2
\\
\nn
&&=
\left(\int_0^1 dx~ q^2(x)\right)^2 =
2\int_0^1 dx~ q^2(x)\int_0^x dy~ q^2(y) \ .
\EEA

%%%%%%%%%%%%%%%%%%%%%%%%%%%%%%%%%%%%%%%%%%%%%%%%%%%%%%%

\subsubsection{Full RSB free energy for disordered BEG around $T_c$}
Computing  the expansion for small $q$ and $\rho\simeq \rho_0$,
 Eq. (\ref{exp}), in the FRSB scheme, yields
\BEA
&&\beta f=\beta f_0-\beta \Kt\left(\rho_0-\frac{T}{J}\right)
-\frac{\Lambda_0}{4}\int_0^1 dx~ q(x)^2
\label{f:C2a}
\\
\nonumber
&&
+\frac{\Lambda_1}{2} r^2-\frac{(\beta \Kt)^3}{6}\rho_0(1-\rho_0)(1-2\rho_0)r^3
\\
\nonumber
&&
-\frac{(\beta\Kt)^4}{24}\rho_0(1-\rho_0)(1-6\rho_0+6\rho_0^2) r^4
\\
\nonumber
&&+\frac{(\beta J)^4}{2}\beta \Kt \rho_0^2(1-\rho_0)
\left[r+\frac{\beta \Kt}{2}(2-3 \rho_0)r^2\right] 
\int_0^1 dx~ q(x)^2
\\
\nn
&&
-\frac{(\beta J)^6}{6}\rho_0^3
\left[1+3\beta\Kt(1-\rho_0) r\right]
\left[3\int_0^1 dx~q(x)\int_0^x dy~q(y)^2 \right.
\\
\nn
&&\hspace*{3 cm}\left.+ \int_0^1 dx~x~q(x)^3\right]
\\
\nonumber
&&+\frac{(\beta J)^8}{8}\rho_0^4
\left[12\int_0^1 dx~q(x)\int_0^x dy~q(y) \int_0^y dz~ q(z)^2
\right.
\\
\nn
&&\hspace*{.5 cm}\left.
+4\int_0^1 dx~q(x)\int_0^x dy~y~q(y)^3
+\int_0^1 dx~x^2 q^4(x)\right]
\\
\nn
&&+\frac{(\beta J)^8}{48}\rho_0^2\left(1-3\rho_0\right)^2
\int_0^1 dx~ q(x)^4
\\
\nn
&&-\frac{(\beta J)^8}{8}\rho_0^3\left(1-3\rho_0\right) 
\left(\int_0^1 dx~ q^2(x)\right)^2 
\\
\nonumber
&&+\frac{(\beta J)^{12}}{1440}\rho_0^2\left(1- 30\rho_0+285\rho_0^2
-900\rho_0^3\right.
\\
\nn
&&\hspace*{ 3 cm}\left.+900\rho_0^4\right)\int_0^1 dx~ q^6(x)
\EEA 

The expansion is valid  along the critical line $\rho=T$ down to the 
tricritical point ($\rho_c,T_c$) given by Eqs. (\ref{f:622b}-\ref{f:622c}).

The last term is introduced for the case $K=0$, i.e. for the
Ghatak-Sherrington model, in which $1-3\rho_0$ goes to zero at the
tricritical point $\rho_c=1/3$ (e.g., decreasing temperature or
density along the second order line). In this case the relevant
quartic term $\int_0^1 dx~ q(x)^4$, responsible for the replica
symmetry breaking, vanishes. This means that, in this
case, the symmetry breaking is weaker than, for instance, in the SK
case.
%\footnote{Because of the importance of the coefficient of such quartic term
%it is sometimes called 
%$\Lambda_{\rm RSB}\equiv (\beta J)^8\rho_0^2(1-3\rho_0)^2/48$} 

%
%Simplifying the Eq. (\ref{exp})
%the  expansion at $K=0$ around the critical point reads:
%\BEQ
%\beta f=\frac{\beta\Kt}{18}+\log\frac{2}{3}-\frac{9}{2}
%\lim_{n\to 0}\frac{1}{n}\Tr{\bf q^4}
%+\frac{729}{40}\lim_{n\to 0}\frac{1}{n}\sum_{ab}q_{ab}^6
%\EEQ

Putting $\rho\to 1$ and $r=0$
in the above expression we obtain the same result
of Ref. [\onlinecite{TAKJPC80}], where the same kind of 
expansion for SK model is performed, up to the forth order.
To allow a straightforward check
we rewrite the above formula in the  way of Ref. [\onlinecite{TAKJPC80}]
\BEA
&&\beta f= \beta f_0
\\
\nonumber
&&-\frac{(\beta J)^2}{4}\left(1-\rho_0^2(\beta J)^2\right)
\int_0^1 dx ~q(x)^2
\\
\nn
&&
-\frac{(\beta J)^6}{6}\rho_0^3
\int_0^1 dx~
\left(x~q(x)^3 + 3 q(x)~\int_0^x dy~q(y)^2  \right)
\\
\nonumber
&&+\frac{(\beta J)^8}{24}\int_0^1 dx~
\left\{
        \left[
                3 \rho_0^4 x^2 
		+\frac{1}{2}\rho_0^2(1-3\rho_0)^2
        \right] 
q(x)^4\right.
\\
\nn
&&\hspace*{2 cm}-2\left. 
        \rho_0^3(1-3\rho_0)~ 
                q(x)^2\int_0^x dy~q(y)^2
\right.
\\
\nn
&&\hspace*{2 cm}+12 \rho_0^4~ q(x)\int_0^x dy~y~q(y)^3
\\
\nn
&&\hspace*{2 cm}\left.
+36 \rho_0^4~ q(x)\int_0^x dy~q(y) \int_0^y dz~ q(z)^2
\right\}
\\
\nn
&&+O(q^6)+O(r)
\EEA 

Variation with respect to $q(x)$ and further differentiation with
respect to $x$ lead to an integral equation that can be reduced to
 \BEQ
\frac{\p_x q(x,\rho)}{q(x,\rho)}=\frac{1}{x}-\frac{6 x
\rho^2}{(1-3\rho)^2+6 x \rho^2} \ 
\label{qdot}
\EEQ
or 
\BEQ
\p_x q(x,\rho)=0
\EEQ

Below the critical temperature the solution is:
\BEQ
q(x,\rho)= \frac{C_1}{(1-3\rho)^2} \frac{x}{\sqrt{(1-3\rho)^2 + 6 \rho^2 x^2}}
\EEQ
\noindent as far as 
  $x\leq x_M(\rho)$, otherwise $q(x,\rho)=q_1$ for $x> x_M(\rho)$.
The value $x_M$ is given by equating  $q_1=q(x_M,\rho)$:
\BEQ
 x_M(\rho) = \frac{q_1 (1-3\rho)^3}{\sqrt{ C_1^2-6\rho^2(1-\rho)^4 q_1^2}} 
\simeq  \frac{q_1 (1-3\rho)^3}{C_1}+
O(q^3)
\label{xM}
\EEQ
Notice that in Eqs. (\ref{qdot}), (\ref{xM}) $\rho=\rho_0+r$ and the 
formulas
have yet to be expanded in $r$.
The above computation can be simplified neglecting the quartic terms in Eq.
(\ref{exp}) which are irrelevant with respect to the RSB, i.e. all but 
the one involving $q_{ab}^4$.
In this case $q(x)$ is simply
\BEQ
q(x)= A(r)~ x 
\EEQ
with
\BEQ
%A(r)\equiv \frac{2 \rho_0}{(1- 3\rho_0)^2}
%\left(+\beta\Kt\frac{(\beta J)^6}{2}\rho_0^3(1-\rho_0) r\right)~ 
A(r)\equiv \frac{2 \rho_0}{(1- 3\rho_0)^2}
+\beta\Kt(\beta J)^6\rho_0^4\frac{(1-\rho_0)}{(1- 3\rho_0)^2}~ r
\EEQ

for $x\leq x_M\sim q_1/A(r)$ and $q(x)=q_1$ for $x>x_M$.
If $r\to 0$ and $\rho_0\to 1$ $A(0)$ reduces to $1/2$ (SK model). 
\cite{SomJPL85}


\begin{thebibliography}{99}


\bibitem{PJPA80} G. Parisi, J. Phys. A {\bf{13}} (1980) L115.

\bibitem{SKPRL75} D. Sherrington, and  S. Kirkpatrick,
Phys. Rev. Lett. {\bf{26}} (1975)  1782.

\bibitem{DPRL80} B. Derrida,  Phys. Rev. Lett. {\bf{45}} (1980) 79.

\bibitem{EAJPF75} S.F. Edwards and  P.W. Anderson, 
 J. Phys. F {\bf{5}} (1975) 965.

\bibitem{GKSPRL85}
D.J. Gross, I. Kanter and h. Sompolinsky, Phys. Rev. Lett.
{\bf 55} (1985) 305.

\bibitem{CSZPB92} A. Crisanti and H.J. Sommers, Z. Phys. B {\bf 87} (1992) 341.


\bibitem{Nie95}
Th. M. Nieuwenhuizen, Phys. Rev. Lett. {\bf 74},  4289 (1995).

\bibitem{CriCuk}
A. Crisanti and S. Ciuchi,
Europhys. Lett. {\bf 9}, 754  (2000).


\bibitem{GolShe85}
P. M. Goldbardt and D. Sherrington, J. Phys. C {\bf 18}, 1923 (1985);
P. M. Goldbardt and Elderfield, J. Phys. C {\bf 18}, L229 (1985);
D. Sherrington, Prog. Theor. Phys. Supp. {\bf 87}, 180 (1986).

\bibitem{GNPB85}
E. Gardner,  Nucl. Phys. B {\bf{257}} (1985) 747.

\bibitem{KWPRA87}
T.R. Kirkpatrick and  P.G.  Wolynes, Phys. Rev. A {\bf{35}} (1987) 3072.

\bibitem{GMNPB84} D.J. Gross, M. Mezard,  Nucl. Phys. B {\bf{240}} (1984) 431.

\bibitem{BCKM98} J. P. Bouchaud, L. Cugliandolo, J. Kurchan and  M.Mezard, in 
{\em Spin Glasses and Random Fields}, A. P. Young, ed.  (World Scientific, 
Singapore, 1998), p. 161.

\bibitem{CHSZPB93}
A. Crisanti, H. Horner, H. J. Sommers, Z. Phys. B {\bf{92}} (1993) 257.

\bibitem{CLPRL02} A. Crisanti and  L. Leuzzi,
Phys. Rev. Lett. {\bf 89} (2002) 237204.

\bibitem{BEGPRA71} H. W. Capel, Physica {\bf{32}}, 966 (1966);
M. Blume, Phys. Rev. {\bf{141}}, 517 (1966);
M. Blume, V.J. Emery and R.B. Griffiths, 
Phys. Rev. A {\bf{4}} (1971) 1071.

\bibitem{HBPRL91}
W. Hoston, A. N. Berker, Phys. Rev. Lett. {\bf{67}} (1991) 1027.

\bibitem{GSJPC77} S. K. Ghatak, D. Sherrington, J. Phys. C: Solid State Phys.
{\bf{10}}, 3149 (1977).

\bibitem{LAJPC82} E.J.S. Lage and J.R.L. de Almeida, J. Phys. C: Solid State 
Phys. {\bf{15}} (1982) L1187.

\bibitem{MSJPC85}P.J. Mottishaw and D. Sherrington, J. Phys. C: 
Solid State Phys. {\bf{18}}  (1985) 5201.



\bibitem{dCYSJPA94}F.A. da Costa, C.S.O. Yokoi and  R.A. Salinas,
J. Phys.A: Math. Gen. {\bf{27}} (1994) 3365.


\bibitem{ANSJPF96}J.J. Arenzon,  M. Nicodemi and  M. Sellitto,  
J. Phys.  I France {\bf{6}} (1996) 1143.

\bibitem{SNAJPF97}M. Sellitto, M. Nicodemi, J.J. Arenzon, J. Phys.  I (France) 
{\bf{7}} (1997) 945.

\bibitem{dCNYJPA97}F.A. da Costa, F. D. Nobre and  C.S.O. Yokoi, J. Phys. A: 
Math. Gen. {\bf{30}} (1997) 2317.

\bibitem{SEPJB99}G.R. Schreiber, Eur. Phys. J. B, {\bf{9}} (1999) 479.

\bibitem{dCAEPJB00}F.A. da Costa, J.M. de Ara\'ujo, Eur. Phys. J. B, {\bf{15}
} (2000) 313.

\bibitem{ANdCJPC00}A. Albino Jr., F. D. Nobre, F. A. da Costa, J. Phys.: 
Cond. Matt. {\bf{12}} (2000) 5713.

\bibitem{AdCNEPJB00}J.M. de Ara\'ujo, F.A. da Costa, F. D. Nobre, Eur. Phys. 
J. B, {\bf{14}} (2000) 661.

\bibitem{CNC1} A. Caiazzo, A. Coniglio and M. Nicodemi, Phys. Rev. E {\bf 66},
046101 (2002).

\bibitem{CNC2} A. Caiazzo, A. Coniglio and M. Nicodemi, {\sf cond-mat 0311442}

\bibitem{CJPIV93} A. Coniglio,  J. Phys. IV France {\bf{3}} (1993) C1-1;
Il Nuovo Cimento D {\bf{16}} (1994) 1027.

\bibitem{dCCPRE01} A. de Candia and A. Coniglio,
Phys. Rev. E {\bf{65}}, 16132 (2001).

\bibitem{NC} M. Nicodemi and A. Coniglio, Phys. Rev. E, {\bf 57}, R39 (1998).

\bibitem{MPV87} M. Mezard, G. Parisi and  M.Virasoro, 
{\em Spin Glass Theory and Beyond} (World Scientific, Singapore, 1987).

\bibitem{Fischer91}K.H. Fischer and J.A. Hertz, {\em Spin Glasses}

\bibitem{DGOJPL81} C. de Dominicis, M. Gabay and  H. Orland, 
J. Phys. Lett. {\bf{42}} (1981) L523.

\bibitem{DGDJPA82} C. de Dominicis, M. Gabay and  B. Duplantier, 
J. Phys. A {\bf{15}} (1982)  L47-L49.

\bibitem{SPRL81} H. Sompolinsky, Phys. Rev. Lett. {\bf{47}} (1981) 935.

\bibitem{SZPRB82} H. Sompolinsky, A. Zippelius, Phys. Rev. B {\bf{25}}
(1982) 6860.

\bibitem{SDJPC84} H. J. Sommers, W. Dupont, J. Phys. C {\bf{17}}  (1984)
 5785.

\bibitem{NJPC87} K. Nemoto, J. Phys. C {\bf 20} (1987) 1325.

\bibitem{CRPRE02} A. Crisanti, T. Rizzo, Phys. Rev. E {\bf{65}} (2002) 
 046137. 

\bibitem{CLPJPA02} A. Crisanti, L. Leuzzi and  G. Parisi,
J. Phys. A: Math. Gen. {\bf 35} (2002) 481.

\bibitem{Orszag71} S. A. Orszag,  {\em Studies in applied mathematics}
        (Cambridge University, Cambridge, 1971, Vol. 4).

\bibitem{Ferziger96}J.H. Ferziger and M. Peri\'c,
{\em Computational Methods for Fluid Dynamics} (springer-Verlag, Berlin, 1996).




\bibitem{MMM1} A. Montanari, M. M\"uller and M. M\'ezard, 
e-print {\sf cond-mat 0307040}.

\bibitem{MMM2} M. M\"uller, A. Montanari and M. M\'ezard, 
e-print {\sf cond-mat 0401139}.


%\bibitem{LPJSP01} L. Leuzzi and  G. Parisi, 
%J. Stat. Phys. {\bf{103}} (2001) 679.


%\bibitem{ESJPC83} D. Elederfield and D. Sherrington, J. Phys. C {\bf 16}
%(1983) L497; L971 (1983)
%\bibitem{LEJPC83} E.J.S Lage and A. Erzan, J. Phys. C {\bf 16} (1983) L873.



\bibitem{MEPL86} P. Mottishaw, Europhys. Lett. {\bf 1} (1986) 409.

%\bibitem{ACNJPA00} J.M. de Ara\`ujo, F.A. da Costa and F.D. Nobre, 
%J. Phys. A {\bf 33} (2000) 1987.





\bibitem{TAPPM77} D.J. Thouless, P.W. Anderson and  R.G. Palmer, 
Phil. Mag. {\bf{35}} (1977) 593.

%\bibitem{CSJPI95} 
%A. Crisanti and  H.J. Sommers, J. Phys. I {\bf{5}} (1995) 805.


%---------------end refs cap 4 ------------------------------------%


%\bibitem{AdCNJPA00}J.M. de Ara\'ujo, F.A. da Costa, F. D. Nobre,
%J. Phys. A {\bf 33} (2000) 1987.

\bibitem{CCL} A. Caiazzo, A. Crisanti and L. Leuzzi, in preparation


\bibitem{TAKJPC80} D.J. Thouless, J.R. de Almeida and J.M. Kosterlitz,
J. Phys. C {\bf 13} (1980) 3271.

\bibitem{SomJPL85} H.-J. Sommers, J. Phys. Lett. {\bf 46} (1985) L-779.

\bibitem{KSSPJETP85}I.Y. Korenblit, E.F. Shender, Sov. Phys. JETP {\bf{62}}, 
1030 (1986).

\end{thebibliography}
\end{document}